\documentclass[aps,prfluids,preprint,superscriptaddress,
showkeys,nofootinbib,longbibliography]{revtex4-2}

\usepackage[colorlinks,bookmarksopen,bookmarksnumbered,citecolor=red,urlcolor=red]{hyperref}
\usepackage{amsmath}
\usepackage{amsfonts}
\usepackage{amssymb}
\usepackage{tikz}
\usetikzlibrary{plotmarks}

\usepackage{caption}
\usepackage{subcaption}
\usepackage{mathtools}
\usepackage{setspace}
\usepackage{graphicx}
\usepackage{pgfplots}

\pgfplotsset{compat=newest}
\pgfplotsset{plot coordinates/math parser=false}
\newlength\figureheight
\newlength\figurewidth

\newcommand{\vinf}{v_\infty}
\newcommand{\kn}{\text{Kn}}
\newcommand{\pie}{\sqrt{\frac{\pi}{2}}}
\newcommand{\eip}{\sqrt{\frac{2}{\pi}}}

\newcommand{\Rho}{\varrho}
\newcommand{\Pres}{\mathcal{P}}
\newcommand{\V}{\mathcal{V}}
\newcommand{\Q}{\mathcal{Q}}
\newcommand{\R}{\mathcal{R}}
\newcommand{\M}{\mathcal{M}}
\newcommand{\Del}{\mathfrak{D}}
\newcommand{\set}[1]{\left \lbrace #1 \right \rbrace}
\newcommand{\Z}{\mathbb{Z}}

\begin{document}
\title{An Efficient Moment Method for Modelling Nanoporous Evaporation}
\author{Thomas C. De Fraja}
\email{tcd2000@hw.ac.uk}
\affiliation{Mathematics Institute, University of Warwick, UK}
\affiliation{School of Engineering, University of Warwick, UK}
\author{Anirudh S. Rana}
\affiliation{Department of Mathematics, Birla Institute of Technology, India}
\author{Ryan Enright}
\affiliation{Nokia Bell Labs, New Jersey, USA}
\author{Laura J. Cooper}
\affiliation{Mathematics Institute, University of Warwick, UK}
\author{Duncan A. Lockerby}
\email{d.lockerby@warwick.ac.uk}
\affiliation{School of Engineering, University of Warwick, UK}
\author{James E. Sprittles}
\email{j.e.sprittles@warwick.ac.uk}
\affiliation{Mathematics Institute, University of Warwick, UK}

%\begin{abstract}
%Evaporation is an effective cooling mechanism widely exploited in the thermal management of modern electronic devices, with thin-film-based nanoporous membrane technologies being a promising candidate for the efficient cooling of microscale and nanoscale devices.
%At these scales, the accuracy of the Navier-Stokes-Fourier equations is reduced, as domain sizes become comparable to the mean free path in the vapour. 
%Here, we investigate the applicability of higher order moment methods, derived from the Boltzmann equation, for simulating evaporation from a nanoporous membrane. 
%Solutions to these are benchmarked against existing solutions of the Boltzmann equation obtained by the method of direct simulation Monte Carlo. 
%The outcome is a simultaneously accurate and computationally cheap method, which can provide simulation-for-design capabilities at the nanoscale.
%\end{abstract}

\begin{abstract}
Thin-film-based nanoporous membrane technologies exploit evaporation to efficiently cool microscale and nanoscale electronic devices.
At these scales, when domain sizes become comparable to the mean free path in the vapour, traditional macroscopic approaches such as the Navier-Stokes-Fourier (NSF) equations become less accurate, and the use of higher-order moment methods is called for.
Two higher-order moment equations are considered; the linearised versions of the Grad 13 and Regularised 13 equations. These are applied to the problem of nanoporous evaporation, and results are compared to the NSF method and the method of direct simulation Monte Carlo (i.e. solutions to the Boltzmann equations).
Linear and non-linear versions of the boundary conditions are examined, with the latter providing improved results, at little additional computational expense, compared to the linear form.
The outcome is a simultaneously accurate and computationally efficient method, which can provide simulation-for-design capabilities at the nanoscale.
\end{abstract}

\keywords{Evaporative cooling, nanoscale flow, gas kinetic effects, higher order moment methods}
\maketitle

\section{Introduction}
Recent developments in technologies, such as laser diodes, power amplifiers, and the upcoming 6G, require more sophisticated thermal-management systems than traditional cooling techniques, as well as more compact solutions. 
For instance gallium nitride based power amplifiers generate hot spots on the micro scale, with heat fluxes of over 1kW/cm$^2$,
temperatures reaching over 180$^\circ$C, and temperature fluctuations of around 40$^\circ$C over
the space of a few microns \citep{GaNtempmeasurements}. 
Such high temperatures and heat fluxes, if unmanaged, lead to a decreased device efficiency and short operating lifetimes.

One promising emerging technology aims to improve heat dissipation capabilities, while also being substantially more compact than traditional cooling techniques. It uses a thin membrane (thickness $\sim$1$\mu$m) covered in nanometre sized pores (pore diameter $\sim$100nm) to transfer heat away via evaporation \citep{Membranepaper,Nanomembrane}. 
An advantage of this cooling mechanism is that the evaporated mass lost is naturally replaced by the capillary action of the liquid (i.e. via wicking).
This saves considerable power and space, as substantial pumping is no longer needed.

Different factors can affect the evaporative capability of nanoporous membranes, including the size of the nanopores \citep{lu2014design}, the thickness of the membrane \citep{lu2014design}, and the working liquid used \citep{Membranepaper}.
The liquid could also have an effect on the shape of the meniscus formed inside the nanopores, which will affect the evaporative surface area, potentially affecting evaporation rates \citep{Nanomembrane}.
Optimising a system over such a large parameter space can be challenging
to explore experimentally, costly in both time and money, and difficult to measure due to the small spatio-temporal scales of interest. 
Computational models offer a cheaper alternative to experiments that enable access to `real time' flow characteristics. 
However, at the nanoscale, additional physics renders the widely used Navier-Stokes-Fourier model inaccurate, motivating the use of both molecular dynamics \citep{MDbook,MDnanoevap} and methods based on the full Boltzmann equation \citep{DSMCbird,FKS,su2020boltz}. Unfortunately, whilst highly accurate, such approaches can be computationally expensive, particularly when simulating device-scale dynamics.

Besides the Boltzmann equation, rarefaction effects that are beyond the resolution
of the NSF system can be predicted by extended macroscopic moment equations
\cite{Strbook,Struchtrup2008HigherorderEI,gu2007computational}.
The moment equations have the favourable form of a set of partial differential equations describing
the evolution of macroscopic quantities, such as mass density, temperature, velocity,
heat flux, stress tensor and so on, defined as moments of the distribution function.
These equations are obtained by an asymptotic reduction of the Boltzmann equation at
different levels of approximation. The moment method was introduced to gas kinetic
theory by Grad \cite{Gradmom}, who expressed the distribution function in terms of Hermite
polynomials. More recently, the regularisation of Grad’s 13-moment (G13) equations
have been obtained by Struchtrup and Torrilhon \cite{Strreg}. The regularised 13-moment
(R13) equations introduce additional terms to the G13 equations that overcome
various deficiencies, such as the prediction of sub-shocks at high Mach number
(Ma $\geqslant$ 1.65). 
Notably, other approaches for reducing the Boltzmann equation have been proposed \cite{fan2016model, koellermeier2014framework}, but these focus on non-linear/hyperbolic regimes beyond the remit of this article.
Importantly, to capture micro/nano-flows, where surface effects can be dominant, Gu and Emerson \cite{gu2007computational} presented a set of wall boundary conditions
for the R13 equations derived from the Maxwell accommodation model, with corrections to this work presented by Torrilhon
and Struchtrup \cite{Struchtrup2008HigherorderEI}.  Subsequently, the R13 equations have been considered for canonical
boundary-value problems, such as planar and cylindrical Couette and Poiseuille flows
\cite{taheri2009macroscopic,taheri2009effects}, transpiration flows and gas flow past
a sphere \cite{R13sphere}, among many others, in one-, two- and three-dimensional numerical
simulations \cite{claydon2017fundamental,rana2013robust, torrilhon2006two,gu2007computational,
r13FEM,r13FEM2,r13FEM3,
claydon2017fundamental,rana2021efficient}.
Building on these successes, we are now able to focus these equations on a specific technological application.

%In this paper, we study an efficient moment method based on the Boltzmann equation, applied to the geometry of a nanoporous membrane.

Section \ref{sec:momeqns} introduces the equations, of which we consider three sets (or `levels' of accuracy/complexity): the Navier-Stokes-Fourier equations (NSF); the Grad 13 equations; and the regularised Grad 13 equations, known as the R13 equations. These are all moment-based approximations of the Boltzmann equation, each more accurate than the previous \citep{torrilhon2016moments}. 
We compare solutions of these equations to computational results obtained by John \emph{et al.} \citep{Benzipaper,Benzipaper2}, who utilised the standard Direct Simulation Monte Carlo (DSMC) method of solving the Boltzmann equation. 

In Sections \ref{sec:linear}, \ref{sec:semilin} and \ref{sec:knudanal}, we present the results of nanoporous membrane simulations based on the moment equations being solved using the finite element method. 
The different models are compared to analytic results and DSMC for both a fully linearised model in Section \ref{sec:linear}, and a model incorporating non-linear boundary conditions in Section \ref{sec:semilin}; the latter providing the most accurate representation of the DSMC simulation of nanoporous evaporation, and a marked improvement over the fully linear model (see e.g. Figure \ref{fig:ap05lnlcomp}).  
Section \ref{sec:knudanal} analyses the effect of scale on the membrane simulations, extending the work done in sections \ref{sec:linear} and \ref{sec:semilin} to higher Knudsen numbers.

Finally, in Section \ref{sec:conc}, conclusions and discussion are presented.

\section{\label{sec:momeqns}The Moment Equations}
The moment equations are derived as approximations of the Boltzmann equation by Grad's moment method \citep{Gradmom}. Details of the derivation are well documented, and are outlined in Appendix \ref{sec:momderiv}. Index notation is used throughout, with indices used for the spatial dimension. The Grad 13 and R13 equations are a set of PDEs in terms of the following moments; the density $\Rho$, the velocity $\V_i$, temperature given in specific energy units $\Theta$, the trace-free and symmetric stress tensor $\Sigma_{ij}$, and the heat flux $\Q_i$. For a formal introduction to these moments, see the Appendix \ref{sec:momderiv} and \citep{Strbook}. This forms a set of 13 moments, which we denote
\[\Phi^{[13]}= \Rho\left \lbrace 1, \mathcal{V}_i, \Theta, \Sigma_{ij}, \mathcal{Q}_i \right \rbrace.
\]
The governing equations for the two 13-moment systems, describing the evolution of the set of moments $\Phi^{[13]}$, consist of the conservation laws;
\begin{equation}\label{eqn:conservationlaws}
\begin{aligned}
D_t \Rho +\Rho \partial_k \mathcal{V}_k=0,\\[5pt]
\Rho D_t \mathcal{V}_i+ \partial_i (\Rho\Theta) + \partial_k \Sigma_{ik}=G_i,\\[5pt]
\frac{3}{2}\Rho D_t \Theta + \Rho\Theta \partial_k \mathcal{V}_k + \partial_k \mathcal{Q}_k+\Sigma_{kl}\partial_k \mathcal{V}_l=0,
\end{aligned}
\end{equation}
with the additional balance equations for stress and heat flux; 
\begin{equation}\label{eqn:stress}
\begin{aligned}
D_t\Sigma_{ij}+\Sigma_{ij}\partial_k \mathcal{V}_k +2\Rho\Theta \partial_{\langle i}\mathcal{V}_{j\rangle} +2\Sigma_{k \langle i} \partial_k \mathcal{V}_{j \rangle}\\[1pt]
+\frac{4}{5}\partial_{\langle i} \mathcal{Q}_{ j \rangle} +\partial_k \mathcal{M}_{ijk} = -\Rho\Theta \frac{\Sigma_{ij}}{\mu},\\
\end{aligned}
\end{equation}
\begin{equation}\label{eqn:heatflux}
\begin{aligned}
D_t \mathcal{Q}_i + \frac 52 \Sigma_{ik} \partial_k \Theta-\Sigma_{ik}\Theta\partial_k\ln \Rho +\Theta \partial_k \Sigma_{ik}+\frac 75 \mathcal{Q}_i \partial_k \mathcal{V}_k\\[1pt]
+\frac{7}{5} \mathcal{Q}_k \partial_k \mathcal{V}_i+\frac 25 \mathcal{Q}_k \partial_i \mathcal{V}_k + \frac 12 \partial_k \mathcal{R}_{ik}+ \frac 16 \partial_i \Del + \mathcal{M}_{ikl}\partial_k \mathcal{V}_l\\[1pt] 
-\frac{\Sigma_{ik}}{\Rho} \partial_l \Sigma_{kl}+\frac 52 \Rho\Theta \partial_i \Theta =-\frac 52\Rho\Theta\frac{\mathcal{Q}_i}{\kappa},
\end{aligned}
\end{equation}
where $\partial_i = \frac{\partial}{\partial x_i}$, $D_t = \partial_t+ v_k \partial_k$ is the material derivative, $G_i$ are external forces, $\mu$ is the dynamic viscosity, and $ \kappa=\frac{15}{4}\mu$ is the heat conductivity. Indices in angular brackets denote the trace-free-symmetric part of a tensor \citep{Strbook}.
 We also assume the gas to be ideal, with the pressure $\Pres=\Rho\Theta$. 

The five-moment NSF system for the moments $\Phi^{[5]}= \Rho\left \lbrace 1, \mathcal{V}_i, \Theta \right \rbrace$ applies closure to equations \eqref{eqn:conservationlaws}, with well-known constitutive laws for stress and heat flux
\begin{align}\label{eqn:NSF}
\Sigma_{ij}= - 2\mu \partial_{\langle i}\mathcal{V}_{j\rangle}, \quad \Q_i = -\kappa \partial_i \Theta,
\end{align}
which give the Navier-Stokes equations and the heat equation (based on Fourier's law). While the NSF system only requires the five moments $\Phi^{[5]}$, it is worth noting that when solving this system numerically, we consider the full 13 moments $\Phi^{[13]}$, in order to retain a first-order system, replacing \eqref{eqn:stress}-\eqref{eqn:heatflux} by \eqref{eqn:NSF}.

{\parindent0pt
\subsubsection*{Closure of the 13 Moment System}
}
The equations \eqref{eqn:conservationlaws}-\eqref{eqn:heatflux} give 13 equations for the 13 moments $\Phi^{[13]}$, but there are the additional moments $\M_{ijk}, \R_{ij}$ and $\Del$.
For the Grad 13 system, the higher moments $\M_{ijk}, \R_{ij}$ and $\Del$ are set to zero \citep{Gradmom,Gradbook,Strbook} in order to close \eqref{eqn:conservationlaws}-\eqref{eqn:heatflux}. 

The R13 system is derived via the order-of-magnitude method as an approximation to the Boltzmann equation from its infinite set of corresponding moment equations \cite{Strknud,Strbook}.
In the order-of-magnitude method, first the leading order of all moments is determined by means of the Chapman-Enskog expansion (in Knudsen number) and then the moment equations are systematically reduced by cancelling terms of higher order. 
The R13 equations are obtained via third order closure, i.e., an order equivalent to Super-Burnett (a model which is derived via a direct Chapman-Enskog expansion of the Boltzmann equation \cite{Strbook}). 
Notably, over the years, a number of different variants of these equations has been suggested with differences occurring in the non-linear terms of the higher moments, which will not concern us here, as we will focus on the linearised form.

The closure for the R13 system is given by
\begin{equation}\label{eqn:R13moments}
\begin{aligned}
2\mu \Theta \partial_{\langle i} \frac{\Sigma_{jk\rangle}}{\Pres} -\frac 43 \frac{\Sigma_{\langle ij}\Q_{k\rangle}}{\Pres}&=-\M_{ijk},\\
\frac{24}{5}\mu\Theta \partial_{\langle i} \frac{\Q_{j\rangle}}{\Pres}-\frac{64}{25}\frac{\Q_{\langle i} \Q_{j\rangle}}{p} -\frac{20}{7}\frac{\Sigma_{k\langle i} \Sigma_{j\rangle k}}{\Rho}&=-\R_{ij},\\
12\mu\Theta \partial_{k} \frac{\Q_{k}}{\Pres}-\frac{56}{5}\frac{\Q_{k} \Q_{k}}{\Pres} -5\frac{\Sigma_{kl} \Sigma_{l k}}{\Rho}&=-\Del.
\end{aligned}
\end{equation}
In a similar vein to the NSF equations, though the R13 equations are a closed set of PDEs for the moments $\Phi^{[13]}$, the full set of 26 moments
\[\Phi^{[26]}= \Rho\left \lbrace 1, \mathcal{V}_i, \Theta, \Sigma_{ij}, \mathcal{Q}_i, \M_{ijk}, \R_{ij}, \Del \right \rbrace
\]
are considered for numerical simulation.

In line with the order-of-magnitude method, at the zeroth order closure the Euler equations are recovered, while the second order closure leads to the NSF equations. The Grad 13 moment equations (for the Maxwellian molecules) were shown to be second order (i.e., Burnett order), and R13 third order (i.e., super-Burnett order). The R26 equations obtained by Gu and Emerson \cite{gu2009high} are fifth order accurate.  

Apart from the accuracy of the macroscopic theories, the equations also differ in their mathematical and physical nature. 
The Grad 13 system, due to their hyperbolic character, produces unphysical sub-shocks for the flows with Mach number, Ma$\geq 1.65$ \cite{torrilhon2004regularized}. 
On the other hand, the NSF and the R13 equations give smooth shock structures for all Mach number \cite{torrilhon2004regularized,timokhin2017different}.
Moreover, the results predicted by the R13 theory are closer to kinetic theory for Ma $\lesssim 5$ \cite{timokhin2017different}.
The Burnett equations are known to show linear unstablites for time-dependent problems \cite{bobylev2006instabilities} while the NSF, Grad 13 and R13 equations are stable \cite{Strknud}.

{\parindent0pt
\subsection{Boundary Conditions}
}
The boundary conditions for the 13-moment systems can be derived by continuity of fluxes of moments of the distribution function at the boundary \cite{StrEvapbc}. The liquid is assumed to be in equilibrium, with temperature $\Theta_L$ and saturation pressure $\Pres_s$. 
The probability that a particle evaporates/condenses at a liquid-vapour portion of the boundary is $\vartheta$. The accommodation coefficient is $\chi$, defined as the probability that a particle is thermalised. 
We consider only 2-dimensional flows, leaving us with the 2-dimensional boundary coordinate system, with normal direction denoted $n$, and tangential direction denoted $t$. The normal coordinate representation of a tensor $T_{i_1,...,i_m}$ is
\[T_{n...n} = T_{i_1...i_m} n_{i_1} ... n_{i_m},
\]
and similarly for tangential and mixed parts. The effective pressure is 
\begin{equation*}
\Pi=\Pres+\frac{1}{2} \Sigma_{n n}-\frac{1}{120} \frac{\Del}{\Theta}-\frac{1}{28} \frac{\R_{n n}}{\Theta}.
\end{equation*}
The resulting boundary conditions \citep{StrEvapbc} are given in (\ref{eqn:evapbc})-(\ref{eqn:highbc3}), with the evaporation mass flow condition
\begin{equation}\label{eqn:evapbc}
\Rho \V_{n}=\frac{\vartheta}{2-\vartheta} \sqrt{\frac{2}{\pi}}\left[\frac{\Pres_s\left(\Theta_{L}\right)}{\sqrt{\Theta_{L}}}-\frac{\Pi}{\sqrt{\Theta}}\right],
\end{equation}
tangential slip condition
\begin{equation}\label{eqn:slipbc}
\begin{aligned}
\Sigma_{t  n}=-\frac{\vartheta+\chi(1-\vartheta)}{2-\vartheta-\chi(1-\vartheta)} \sqrt{\frac{2}{\pi \Theta}}\left[\Pi \V_{t }+\frac{1}{5} \Q_{t }+\frac{1}{2} \M_{t  n n}\right]-\Rho \V_{n} \V_{t },
\end{aligned}
\end{equation}
and the normal heat transfer condition
\begin{equation}\label{eqn:tempjumpbc}
\begin{aligned}
\Q_{n}=&-\frac{\vartheta+\chi(1-\vartheta)}{2-\vartheta-\chi(1-\vartheta)} \sqrt{\frac{2}{\pi \Theta}}
\left[2 \Pi\left(\Theta-\Theta_{L}\right)-\frac{\Pi}{2} \V_{t}^{2}+\frac{1}{2} \Theta \Sigma_{n n}+\frac{\Del}{15}+\frac{5}{28} \R_{n n}\right] \\
&+\left[\frac{1}{2}\left(\V_{t}^{2}-\Theta_{L}\right)-\frac{5}{2}\left(\Theta-\Theta_{L}\right)\right] \Rho \V_{n}.
\end{aligned}
\end{equation}
The R13 equations require additional boundary conditions on top of \eqref{eqn:evapbc}-\eqref{eqn:tempjumpbc}, obtained by considering higher moment fluxes \citep{StrEvapbc}, and are given by
\begin{equation}
\begin{aligned}\label{eqn:highbc1}
\M_{n n n}=& \frac{\vartheta+\chi(1-\vartheta)}{2-\vartheta-\chi(1-\vartheta)} \sqrt{\frac{2}{\pi \Theta}} \left[\frac{2}{5} \Pi\left(\Theta-\Theta_{L}\right)-\frac{3}{5} \Pi \V_{t}^{2}-\frac{7}{5} \Theta \Sigma_{n n}+\frac{\Del}{75}-\frac{1}{14} \R_{n n}\right]\\
 &-\frac{2}{5}\left[\Theta_{L}+\frac{3}{2} \V_{t}^{2}\right] \Rho \V_{n},
\end{aligned}
\end{equation}
\begin{equation}\label{eqn:highbc2}
\begin{aligned}
\M_{t  t  n}=&-\frac{\vartheta+\chi(1-\vartheta)}{2-\vartheta-\chi(1-\vartheta)} \sqrt{\frac{2}{\pi \Theta}} \\
& \times\left[\Theta \Sigma_{t  t }-\Pi \V_{t }^2 +\frac{\R_{t  t }}{14}+\frac{1}{5} \Pi\left(\Theta-\Theta_{L}\right)+\frac{1}{5} \Pi \V_{t}^{2}-\frac{1}{5} \Theta \Sigma_{n n}+\frac{\Del}{150}\right] \\
&+\frac{1}{5}\left[ 4\V_{t }^2+\Theta_{L}\right]\Rho \V_{n},
\end{aligned}
\end{equation}
\begin{equation}\label{eqn:highbc3}
\begin{aligned}
\R_{t  n}=& \frac{\vartheta+\chi(1-\vartheta)}{2-\vartheta-\chi(1-\vartheta)} \sqrt{\frac{2}{\pi \Theta}} \\
& \times\left[\Pi \Theta \V_{t }-\frac{11}{5} \Theta \Q_{t }-\frac{1}{2} \Theta \M_{t  n n}-\Pi \V_{t }^3+6 \Pi \V_{t }\left(\Theta-\Theta_{L}\right)\right] \\
&+\left[7\left(\Theta-\Theta_{L}\right)+\Theta_{L}-\V_{t } ^2\right] \Rho \V_{t } \V_{n}.
\end{aligned}
\end{equation}

Notably, the mass-flux condition (\ref{eqn:evapbc}) is the generalisation of the classical Hertz-Knudsen-Schrage law to the higher moment equations, with the vapour pressure $\mathcal{P}$ being replaced by the effective vapour pressure $\Pi$, and the pre-factor being twice as large \citep{Bondbalance}. 
The condition (\ref{eqn:slipbc}) can be seen as a generalised slip condition, relating the tangential velocity $\V_t$ with the shear stress $\Sigma_{tn}$, c.f. the Navier slip model \citep{navier1823memoire,wangslip}. 
Similarly, (\ref{eqn:tempjumpbc}) relates the temperature jump $\Theta-\Theta_L$ with the heat flux $\Q_n$, extending the temperature continuity condition at the boundary. 
Equations (\ref{eqn:highbc1})-(\ref{eqn:highbc3}) are interface conditions for the higher moments.
For non-evaporating surfaces (e.g. at a solid boundary), where $\vartheta=0$, equation \eqref{eqn:evapbc} reduces to the impermeability condition $\V_n=0$, and reproduces the well known jump coefficient $\frac{\chi}{2-\chi}$ for the remaining equations \eqref{eqn:slipbc}-\eqref{eqn:highbc3}, and removes additional contributions of mass flux, giving the vapour-solid R13 boundary conditions \citep{R13microbc}. 

In dealing with fluid flow through micro-devices, one is faced with the question of which model to use, which boundary conditions to apply and how to proceed to obtain solutions to the problem at hand. Surface effects dominate in small devices: the surface-to-volume ration for a device with character length of 1 m is 1 m$^{-1}$, while that for microelectromechanical systems (MEMS) device of size 1 $\mu$m is $10^6$ m$^{-1}$. The million fold increase in surface area relative to the mass of the minute device substantially affects the transport of mass, momentum and energy through the surface \cite{gad1999fluid}.
%In rarefied gas flows, boundary phenomena, such as temperature jump and velocity slip, can have a dominant influence on the flow field. 
As such, proper modelling of the boundary conditions is essential, and it is often the case, particularly when approaching the transition regime, that much of the extra behaviour captured by the higher moment equations is dominated by modifications to the boundary conditions \citep{ranaPRL}, as will be confirmed in Section \ref{sec:nonlinan}.
Still, gas rarefaction also leads to novel bulk phenomena, such as the Knudsen layer, a region of few mean free path lengths thickness where non-equilibrium effects dominates. A detailed examination of the moment models shows that in higher order moment equations the Knudsen layer  appears as superpositions of several exponential layers with different coefficients in their exponents \cite{rana2021efficient,taheri2009macroscopic}. Notably, the NSF and Grad 13 theories cannot capture the Knudsen layer, whilst the R13 theory does, with more moments (e.g. R26) giving a better description of it 
\cite{taheri2009macroscopic,gu2009kramers}.
 
Over the years, a variety of methods have been devised for the simulation of different moment-based systems \cite{cai2018numerical,koellermeier2017numerical,
koellermeier2018two} and, in particular, the R13 equations have been successfully applied to a range of boundary value problems \cite{taheri2009macroscopic,claydon2017fundamental}. Different numerical schemes have been developed for them, for example finite difference \cite{rana2013robust}, finite volume \cite{torrilhon2006two,gu2007computational}, finite element method \cite{r13FEM,r13FEM2,r13FEM3}, and mesh-free methods such as the method of fundamental solutions \cite{claydon2017fundamental,rana2021efficient,
lockerby2016fundamental}.

\mbox{}

{\parindent0pt
\subsection{Linearisation}
}
We simplify the above equations and boundary conditions by linearising about a homogeneous state described by the temperature of the liquid $\Theta_L$ and saturation pressure $\Pres_s$. The saturation density $ \Rho_s$ is such that $\Pres_s=\Rho_s \Theta_L $. Therefore, we shall consider small perturbations from the equilibrium and introduce characteristic scales (i.e. non-dimensionalise) such that \citep{Strreg}
\begin{equation}\label{eqn:linearisationparams}
\begin{aligned}
\Rho &=\Rho_{s}(1+\rho), \quad  \Theta=\Theta_{L}(1+\theta), \quad \Pres=\Pres_{s} (1+\rho+\theta), \\
\mathcal{V}_{i}&=\sqrt{\Theta_{L}}v_i, \quad \Sigma_{i j} =\Rho_{s} \Theta_{L} \sigma_{i j},  \quad \Q_{i}=\Rho_{s} \sqrt{\Theta_{L}} q_{i},\\
\Del&=\Rho_{s} \Theta_{L}^{2} \Delta, \quad \R_{i j}=\Rho_{s} \Theta_{L}^{2} R_{i j},
\,\,\, \M_{i j k} =\Rho_{s} \sqrt{\Theta_{L}}^3 m_{i j k}.
\end{aligned}
\end{equation}
The linear (dimensionless) moments \[ \varphi^{[26]}=\rho\left \lbrace 1, v_i, \theta, \sigma_{ij}, q_i, \Delta, R_{ij},m_{ijk} \right \rbrace \] are taken to be small, thus only linear terms in these will appear in the equations. In this regime, the ideal gas law becomes $p= \rho+\theta.$

In Sections \ref{sec:linear}-\ref{sec:knudanal}, we assess parameters in comparison to DSMC results, for which we require the dimensional moments $\Phi^{[26]}$. We note that the equations are solved both analytically and numerically in terms of the linear non-dimensional moments $\varphi^{[26]}$, after which, for comparison to DSMC, they are plotted in terms of the dimensional moments $\Phi^{[26]}$.

We non-dimensionalise the dependent variables;
\begin{align*}
x_{i}=L \hat{x}_{i}, \quad t=\frac{L}{\sqrt{\Theta_{L}}} \hat{t},
\end{align*}
where $L$ is the characteristic length scale. The hat notation for the dimensionless-dependent variables is henceforth dropped. The Knudsen number is defined as 
\begin{align*}
\mathrm{Kn}=\frac{\mu \sqrt{\Theta_{L}}}{\Pres_{s} L}=\frac{\lambda}{L}
\end{align*}
where $\lambda$ is the mean free path of a particle. Therefore, in this setup, variations in length (e.g. pore diameter) are obtained by varying the Knudsen number.

The saturation pressure and liquid temperature can also be considered as perturbations from their spatially homogeneous equilibrium, with
\begin{align*}
\Pres_s = \Pres_0 (1+p_s), &&\Theta_L = \Theta_0(1+\theta_L).
\end{align*} 
For the most part we will consider constant liquid temperature and saturation pressure, thus $p_s=\theta_L=0$. However, these terms will be left in the boundary conditions to more easily indicate pressure and temperature jumps.
We also consider stationary solutions, i.e. $\partial_t=0$.

Substitution into (\ref{eqn:conservationlaws})-(\ref{eqn:highbc3}) yields the dimensionless and linearised conservation laws
\begin{equation}\label{eqn:linconservationlaws}\tag{\ref*{eqn:conservationlaws}${}^\prime$}
\begin{aligned}
\partial_k v_k &=0,\\
\partial_i \theta +\partial_i \rho +\partial_k \sigma_{ik} &= G_i,\\
\partial_k q_k &=0,
\end{aligned}
\end{equation}
and the expressions for stress and heat flux
\begin{equation}\label{eqn:linstress}\tag{\ref*{eqn:stress}${}^\prime$}
\begin{aligned}
-\frac{\sigma_{ij}}{\kn} = \frac{4}{5} \partial_{\langle i} q_{j\rangle}+ 2 \partial_{\langle i} v_{j\rangle} + \partial_k m_{ijk},
\end{aligned}
\end{equation}
\begin{equation}\label{eqn:linheatflux}\tag{\ref*{eqn:heatflux}${}^\prime$}
\begin{aligned}
-\frac 23 \frac{q_i}{\kn} = \frac{5}{2} \partial_i \theta+ \partial_k \sigma_{ik}  + \frac{1}{2}\partial_k R_{ik} + \frac{1}{6} \partial_i \Delta.
\end{aligned}
\end{equation}
The linearised NSF expressions for stress and heat flux are
\begin{align}\label{eqn:NSFlin}\tag{\ref*{eqn:NSF}${}^\prime$}
\sigma_{ij}= - 2\kn \partial_{\langle i}v_{j\rangle}, \quad q_i = -\frac{15}{4} \kn \partial_i \theta.
\end{align}
The higher moments $m_{ijk}, R_{ij}, \Delta$ all vanish for the Grad 13 system, while the closure of the R13 system is given by
\begin{equation}\label{eqn:linR13moments}\tag{\ref*{eqn:R13moments}${}^\prime$}
\begin{aligned}
-\frac{m_{ijk}}{\kn} = 2\partial_{\langle i} \sigma_{jk\rangle}, &&
-\frac{R_{ij}}{\kn} = \frac{24}{5}\partial_{\langle i} q_{j\rangle}, &&
-\Delta =  0
\end{aligned}
\end{equation}
The linearised boundary conditions are given by
\begin{equation}\label{eqn:linevapbc}
\tag{\ref*{eqn:evapbc}${}^\prime$}
v_{n}=\frac{\vartheta}{2-\vartheta} \sqrt{\frac{2}{\pi}}\left(p_s-\Pi-\frac{1}{2}\theta_L\right),
\end{equation}
\begin{equation}\label{eqn:linslipbc}
\tag{\ref*{eqn:slipbc}${}^\prime$}
\begin{aligned}
\sigma_{t  n}=&-\tilde{\chi} \left(v_{t }+\frac{1}{5} q_{t }+\frac{1}{2} m_{t  n n}\right),
\end{aligned}
\end{equation}
\begin{equation}\label{eqn:lintempjumpbc}
\tag{\ref*{eqn:tempjumpbc}${}^\prime$}
\begin{aligned}
q_{n}=-\tilde{\chi} &\left(2 \left(\theta-\theta_{L}\right)+\frac{1}{2}\sigma_{n n}+\frac{5}{28} R_{n n}\right)-\frac{1}{2}v_n,
\end{aligned}
\end{equation}
\begin{equation}\label{eqn:linhighbc1}
\tag{\ref*{eqn:highbc1}${}^\prime$}
\begin{aligned}
m_{n n n}=\tilde{\chi}& \left(\frac{2}{5}\left(\theta-\theta_{L}\right)-\frac{7}{5} \sigma_{n n}-\frac{1}{14} R_{n n}\right)-\frac{2}{5}v_n,
\end{aligned}
\end{equation}
\begin{equation}\label{eqn:linhighbc2}
\tag{\ref*{eqn:highbc2}${}^\prime$}
\begin{aligned}
m_{t  t  n}=-\tilde{\chi}& \left( \sigma_{t  t }+\frac{R_{t  t }}{14} +\frac{1}{5} \left(\theta-\theta_{L}\right)-\frac{1}{5} \sigma_{n n}\right) +\frac{1}{5} v_n,
\end{aligned}
\end{equation}
\begin{equation}\label{eqn:linhighbc3}
\tag{\ref*{eqn:highbc3}${}^\prime$}
\begin{aligned}
R_{t  n}= \tilde{\chi}&\left(v_t-\frac{11}{5} q_{t }-\frac{1}{2} m_{t  n n}\right),
\end{aligned}
\end{equation}
where
\begin{align*}
\tilde{\chi} = \frac{\vartheta+\chi(1-\vartheta)}{2-\vartheta-\chi(1-\vartheta)} \sqrt{\frac{2}{\pi }},
\end{align*}
and the effective pressure $\Pi$ is \[\Pi =\rho +\frac{1}{2}\theta+\frac{1}{2} \sigma_{n n}-\frac{1}{28} R_{n n}.\]
For all cases considered in this work, fully diffuse molecular re-emissions are
considered at the interface (i.e. $\chi$ = 1), as this is used for the DSMC results \citep{Benzipaper,Benzipaper2}. The evaporation coefficient is zero at a solid-vapour interface, and we assume perfect evaporation at a liquid-vapour interface, i.e. $\vartheta=1$.

For comparison, we solve the Navier-Stokes-Fourier equations with conventional boundary conditions; that is, (\ref{eqn:linslipbc}) is replaced by $v_t=0$ (no slip), and (\ref{eqn:linheatflux}) is replaced by $\theta=\theta_L=0$ (no temperature jump). The evaporation condition (\ref{eqn:linevapbc}) remains.

For the avoidance of doubt, we solve the following systems: NSF, with conservation laws \eqref{eqn:linconservationlaws} and closure \eqref{eqn:NSFlin}, with the evaporation boundary condition \eqref{eqn:linevapbc} accompanied by no slip and no temperature jump; Grad 13, with conservation laws \eqref{eqn:linconservationlaws}, balance equations \eqref{eqn:linstress}-\eqref{eqn:linheatflux} and closure $m_{ijk}=0, R_{ij}=0, \Delta=0$, with boundary conditions \eqref{eqn:linevapbc}-\eqref{eqn:lintempjumpbc}; R13, with conservation laws \eqref{eqn:linconservationlaws}, balance equations \eqref{eqn:linstress}-\eqref{eqn:linheatflux} and closure \eqref{eqn:linR13moments}, with boundary conditions \eqref{eqn:linevapbc}-\eqref{eqn:linhighbc3}. 

\mbox{}

\section{\label{sec:probform}Problem Formulation for Nanoporous Evaporation}
The nanoporous membrane is assumed to be two-dimensional in order to allow comparison with the DMSC results of John \emph{et al.} \cite{Benzipaper,Benzipaper2}. 
It is also necessary to validate the moment equations for two-dimensional flows before modelling a three-dimensional nanoporous membrane, where the computational tractability of the model becomes more important. 

\begin{figure}
\centering
\includegraphics[trim={0 0cm 0 0cm}]{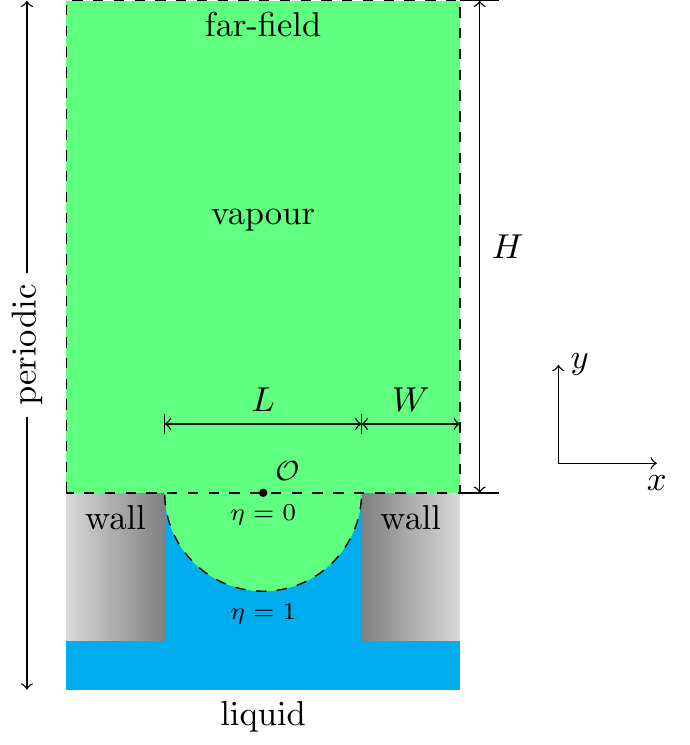}
\caption{\label{fig:geometry}Geometry of a single nanopore. 
Two cases of meniscus curvature $\eta$ are shown, with $\eta=0$ corresponding to the flat meniscus, and $\eta=1$ corresponding to the semi-circular meniscus. 
The porosity shown here is $\phi=0.5$, though porosity can take any value between 0 and 1.}
\end{figure}

The geometry of a single pore of the nanoporous membrane is shown in Figure \ref{fig:geometry}, with this work focussing on the unconventional modelling of the vapour flow, where we can compare to DSMC, with future work coupling this to the  liquid's (conventional) dynamics. The vapour part (green) is the domain of computation where we solve the moment equations. 
It is a rectangle, with a circular segment attached when considering a curved meniscus.
The origin is marked $\mathcal{O}$, and the base of the rectangle is the interface with the nanoporous membrane at $y=0,$ and the top of the rectangle represents a far-field at $y=H$. 
The height $H$ of the rectangle is set to be large enough so that parameters decay sufficiently to an equilibrium before reaching the far-field. 
Parameters at the far-field will be denoted $$\psi_\infty:=\psi(x,y=H).$$
We assume the geometry to be periodic in $x$, with the length of periodicity -- the width of the rectangle -- being the combined width of the meniscus, $L$, and the width of the walls\footnote{The physical width of the wall would be $2W$ when the wall from the adjacent pore is considered.}, $W$, either side.
The porosity $\phi$ is defined as the proportion of length consisting of the meniscus, so $\phi=L/(2W+L)$.
We take $L$ as the characteristic length scale, so that in dimensionless parameters, we have $L=1$.
A curved meniscus is modelled by adding a segment of a circle whose corners are at $x=-\frac{1}{2}$ and $x=\frac{1}{2}$. The meniscus shape is controlled by varying the radius of the circle, $R$. The curvature is then $\eta=\frac{1}{2R}$, so that a flat meniscus corresponds to $\eta=0$, and a semicircular meniscus corresponds to $\eta=1$.

In accordance with the DSMC approach \cite{Benzipaper,Benzipaper2}, we impose a constant dimensionless saturation pressure $p_s$ and liquid temperature $\theta_L$ at the liquid-vapour interface. 
At the far-field, we have three unknown parameters, $v_{i,\infty}, \rho_\infty, \theta_\infty$. We are free to set one of these \citep{ytrehusevap}; we set the velocity field $v_{i,\infty}=\lbrace 0,v_\infty \rbrace$, and solve for the other two.  We also consider the process as adiabatic, so that $q_{i,\infty}=0$.

The linearised 13 moment equations (\ref{eqn:linconservationlaws})-(\ref{eqn:linR13moments}) are solved on the interior of the domain. 
At the liquid-vapour interface, we require the boundary conditions (\ref{eqn:linevapbc})-(\ref{eqn:lintempjumpbc}) for the Grad 13 equations, and additionally the boundary conditions (\ref{eqn:linhighbc1})-(\ref{eqn:linhighbc3}) for the R13 equations. 
At the solid-vapour interface, we have $\vartheta=0$, so that (\ref{eqn:linevapbc}) becomes $v_n=0$. 

The equations are solved using COMSOL Multiphysics v5.6, a commercial finite-element-method (FEM) software (COMSOL Inc, Sweden) \citep{COMSOL}. Guided by standard approaches to incompressible Navier-Stokes simulation \citep{GreshoFEM}, we use the Taylor-Hood elements for momentum, with linear element discretizations for density and quadratic elements for velocity (and stress). Building on this methodology for FEM we use linear elements for the temperature, and heat flux and higher order moments have quadratic elements.
In earlier works \cite{rana2016thermodynamically,Strknud,sarna2018stable}, the derivation of entropy consistent boundary conditions were presented for the linearised moment equations. 	 
For the problems considered in this paper we adopted the boundary conditions proposed in \cite{StrEvapbc}, and while work has been done recently to obtain stabilised R13 FEM solvers for this system \cite{r13FEM,r13FEM2,r13FEM3}, no instabilities were encountered here.

The height is set to be $H = 5+W + 5\kn $, as this was found to always leave enough room for parameters to reach free-stream equilibrium conditions. 
As an example, when $\phi=0.5$ and $\kn=0.05$, the mesh consists of 12,369 triangular domain elements and 580 boundary elements for the flat meniscus case, and 14,115 triangular domain elements and 637 boundary elements for the semicircular meniscus case. These meshes are shown in Figure \ref{fig:meshes} in Appendix \ref{sec:meshpic}.

The DSMC results \citep{Benzipaper,Benzipaper2} have been carried out by John et al. using SPARTA \citep{SPARTA}, which is a highly scalable parallel open-source DSMC code \citep{DSMCopensource,DSMCopensource2}. The gas was assumed to be 
argon and the variable hard sphere (VHS) model was employed. 
It was shown that the DSMC method can reproduce results of the Boltzmann equation, and so it serves as a good benchmark for the moment method \citep{DSMCequiv1,DSMCequiv2,DSMCequiv3}.

\section{\label{sec:linear}Results - Linear Equations }
Simulations allow us to study the dynamics of the evaporative process for various membrane porosities and meniscus shapes. 
To do so, we evaluate the free-stream parameters $\Rho_\infty, \Theta_\infty$ and mass flux $J_\infty = \V_\infty \Rho_\infty $ with respect to the free-stream Mach number, $\mathrm{Ma}_\infty=\V_\infty/\sqrt{\gamma\Theta_\infty}=v_\infty/\sqrt{\gamma\left (1+\theta_\infty\right )},$ where $\gamma=\frac 53$ is the specific heat ratio for a monatomic gas. 
Our choice of linearisation restricts us to small Mach numbers.

{\parindent0pt
\subsubsection*{Analytic Results for a Planar Evaporative Interface (1D Case)}
}

The limiting case $W\to 0$ with a flat meniscus corresponds a planar evaporative interface. 
Analytic (one-dimensional) results have previously been obtained in this limit by various moment methods \cite{StrEvapbc,1DMF}, and are useful as a benchmark for our computational results, particularly given the complexity of the R13 PDE system, even once linearised. Here, we solve all three systems analytically, first finding the general solution to the R13 equations, of which the Grad 13 and NSF are seen as special cases that can be obtained upon setting various terms to zero. The boundary condition system derived is then solved to find far-field parameter expressions for each set of equations. The R13 boundary condition system was previously obtained by Struchtrup \emph{et al.} \cite{StrEvapbc}.

Here, we make the assumption that all parameters do not have $x$ or $z$ dependence, yielding a one-dimensional system which is straightforward to solve, as considered by Struchtrup \emph{et al.} \cite{StrEvapbc}. The conservation equations (\ref{eqn:linconservationlaws}) are the same for all three models, and reduce to
\begin{align*}
\partial_y v_2 =0, \quad \partial_y \sigma_{12}=0,& \quad  \partial_y q_2=0,\\
\partial_y(\rho+\theta+ \sigma_{22}) &=0.
\end{align*}
The R13 expressions for stress are
\begin{align}\label{eqn:str1D}
\begin{aligned}
\sigma_{11} &= \underline{\frac 23 \kn^2 \partial_{yy}\sigma_{11}}-\underline{\frac{4}{15} \kn^2 \partial_{yy}\sigma_{22}},\\
\sigma_{12} &= \frac 25\partial_y   q_1  + \overline{\partial_y v_1} ,\\
\sigma_{22} &= \underline{\frac 65 \kn^2 \partial_{yy}\sigma_{22}},
\end{aligned}
\end{align}
and the expressions for heat flux,
\begin{align}\label{eqn:flux1D}
\begin{aligned}
q_1 &=  \underline{\frac 95 \kn^2 \partial_{yy} q_1},\\
q_2 &= -\overline{\frac{15}{4} \kn \partial_y \theta } - \frac{3}{2} \kn \partial_y \sigma_{22}.
\end{aligned}
\end{align}

In (\ref{eqn:str1D}) and (\ref{eqn:flux1D}), terms underlined indicate those that only appear in the R13 system, while terms overlined indicate the only (derivative) terms that appear in the NSF system.

Integration of the above gives the general solution
\begin{align}\label{eqn:gensol}
\begin{aligned}
v_2&=C_1, \qquad v_1 + \frac 25 q_1= -\frac{C_2}{\kn} y + C_5,\\
\rho + \theta + \sigma_{22} &= C_3, \qquad
\frac{5}{2} \theta + \sigma_{22} = -\frac 23\frac{C_4}{\kn}y + C_6\\
q_2&=C_4,\qquad
q_1 = C_9 \exp \left (-\frac{\sqrt{5}}{3} \frac{y}{\kn} \right ),\\
\sigma_{11} = C_7 \exp & \left ( -\sqrt{\frac{3}{2}} \frac{y}{\kn} \right ) - \frac{1}{2}C_8 \exp \left ( -\sqrt{\frac{5}{6}} \frac{y}{\kn} \right ),\\
\sigma_{12}&=C_2,\qquad\sigma_{22} = C_8 \exp \left ( -\sqrt{\frac{5}{6}} \frac{y}{\kn} \right )
. \\
\end{aligned}
\end{align}
The positive exponential terms have been neglected since we may take $y\to \infty$, and require that parameters are bounded in the far-field. This can also be used to conclude that $C_2 = C_4 =0$. The Grad 13 and NSF systems have $C_{7,8,9}=0$.

At the far field, $y\to\infty$, we have $v_{i,\infty}= \left\lbrace 0,v_\infty \right\rbrace, q_{i,\infty}= \left\lbrace 0,0 \right\rbrace$. Thus $C_1= v_\infty$ and $C_5=0$. Therefore, for NSF and Grad 13 we have $v_1=0$, as expected. 

With far-field notation, we find that $ \theta_\infty =\frac{2}{5}C_6, \rho_\infty = C_3 - \theta_\infty, $ i.e.
\begin{align*}
\theta &= \theta_\infty-\frac{2}{5}C_8  \exp \left( -\sqrt{\frac{5}{6}} \frac{y}{\kn}\right),\\
\rho &= \rho_\infty-\frac{3}{5}C_8  \exp \left ( -\sqrt{\frac{5}{6}} \frac{y}{\kn} \right ).
\end{align*}
At the meniscus, $y=0$, we have the boundary conditions (\ref{eqn:linevapbc})-(\ref{eqn:linhighbc3}). For R13 there are 5 unknowns, $\theta_\infty, \rho_\infty, C_7, C_8, C_9,$ whereas for NSF and Grad 13 there are 2, $\theta_\infty$ and $\rho_\infty.$
The NSF no temperature jump condition immediately gives $\theta_\infty=\theta_L = 0$ for the NSF system.

For the R13 system, equation (\ref{eqn:linslipbc}) gives $C_9=0$, i.e. $v_1=q_1=0$, as expected. The constant $C_7$ only appears in equation (\ref{eqn:linhighbc2}), and so can easily be expressed in terms of $\theta_\infty, \rho_\infty$ and $C_8$.

Boundary condition (\ref{eqn:linevapbc}) is needed for the NSF system to obtain $\rho_\infty$; (\ref{eqn:linevapbc}) and (\ref{eqn:lintempjumpbc}) are needed for the Grad 13 system to obtain $\theta_\infty, \rho_\infty$; (\ref{eqn:linevapbc}), (\ref{eqn:lintempjumpbc}) and (\ref{eqn:linhighbc1}) are required to solve for $\theta_\infty, \rho_\infty, C_8$ in the R13 system. In all cases, the boundary system may be written as 
\begin{align}\label{eqn:bcsystem}\tag{$*$} 
\mathbf{b} = A \mathbf{U}
\end{align}
with $\mathbf{U}^{[\text{NSF}]} = \left\lbrace \rho_\infty \right\rbrace^T, \mathbf{U}^{[\text{G}13]} = \left\lbrace \rho_\infty, \theta_\infty \right\rbrace^T, \mathbf{U}^{[\text{R}13]} = \left\lbrace \rho_\infty, \theta_\infty, C_8 \right\rbrace^T,$ 
\begin{align*}
\mathbf{b}^{[\text{NSF}]} = \left [\begin{matrix}
\vinf
\end{matrix}\right ], &&
\mathbf{b}^{[\text{G}13]} = \left [\begin{matrix}
\vinf\\
\frac{\vinf}{2}
\end{matrix}\right ], && 
\mathbf{b}^{[\text{R}13]} = \left [\begin{matrix}
\vinf\\
\frac{\vinf}{2}\\[3pt]
\frac{2\vinf}{5}
\end{matrix}\right ],
\end{align*}
\begin{align*}
A^{[\mathrm{R}13]} = \left [\begin{matrix}
-\eip & -\frac{1}{\sqrt{2\pi}} &  \frac{3}{5\sqrt{2\pi}}  \\
0 & - \frac{4}{\sqrt{2\pi}} &  \frac{3}{5\sqrt{2\pi}} \\
0 & \frac{4}{5\sqrt{2\pi}} & -\frac{39}{25} \eip -\frac{ \sqrt{30}}{5}
\end{matrix} \right ],
\end{align*}
$A^{[\text{G}13]}$ is the upper left $2\times 2$ sub-matrix of $A^{[\text{R}13]}$, and $A^{[\text{NSF}]}$ is the upper left entry of $A^{[\text{R}13]}$.

Struchtrup \emph{et al.} \cite{StrEvapbc} previously arrived at the above boundary-condition system for the R13 equations, however explicit expressions for the solved parameters were not given. Here, we solve (\ref{eqn:bcsystem}) to give explicit expressions for discussion: the Grad 13 solution has
\begin{align*}
\theta_\infty^{[\text{R}13]} = -\frac{1}{4}\pie\vinf, && 
\rho_\infty ^{[\text{R}13]} = -\frac{7}{8}\pie\vinf,
\end{align*}
and R13 has
\begin{align*}
\theta_\infty^{[\text{R}13]}&= -K\left (2\sqrt{30}\pi +18 \sqrt{2 \pi }\right ) \vinf, \\ 
\rho_\infty ^{[\text{R}13]} &= -K \left (7   \sqrt{30}\pi +57 \sqrt{2 \pi } \right )\vinf,\\
C_8 &= - 10K \sqrt{2\pi} \vinf
\end{align*}
where $K=\frac{1}{16 \sqrt{15 \pi }+120}$. The temperature expression for Grad 13, $\theta_\infty ^{[\text{G}13]}$, is in-line with the analytic expression for temperature obtained by Labuntsov and Kryukov \cite{1DMF}, obtained via an alternate approximation of the Boltzmann equation.

For classical NSF we have
\begin{align*}
\theta_\infty^{[\text{NSF}]} = 0, && \rho_\infty ^{[\text{NSF}]}= - \sqrt{\frac{\pi}{2}} v_\infty.
\end{align*}

The expressions \eqref{eqn:gensol} give insight into some of the differences between the three moment systems. All solutions for the NSF and Grad 13 systems are spatially homogeneous, whereas for the R13 system there is an exponential decay on the order of the Knudsen number.
This represents the fact that the R13 system is able to capture Knudsen-boundary-layer effects \citep{Strreg,torrilhon2016moments}, which typically extend a few mean free paths, $\lambda$, from the boundary \citep{CercignaniRGD}.
This phenomenon also manifests in the two-dimensional flows seen in the numerical results, and is discussed in Section \ref{sec:knudanal}.

The far-field solutions for all systems, however, are independent of Knudsen number, as the characteristic length $L$ is irrelevant in the evaluation of far-field parameters when considering an infinitely long plane of liquid and a one-dimensional flow. In Section \ref{sec:knudanal}, we see that far-field solutions are dependent on Knudsen number when two-dimensionality is introduced.

\begin{figure}
\includegraphics[scale=1.289, trim={0 0cm 0 0cm}]{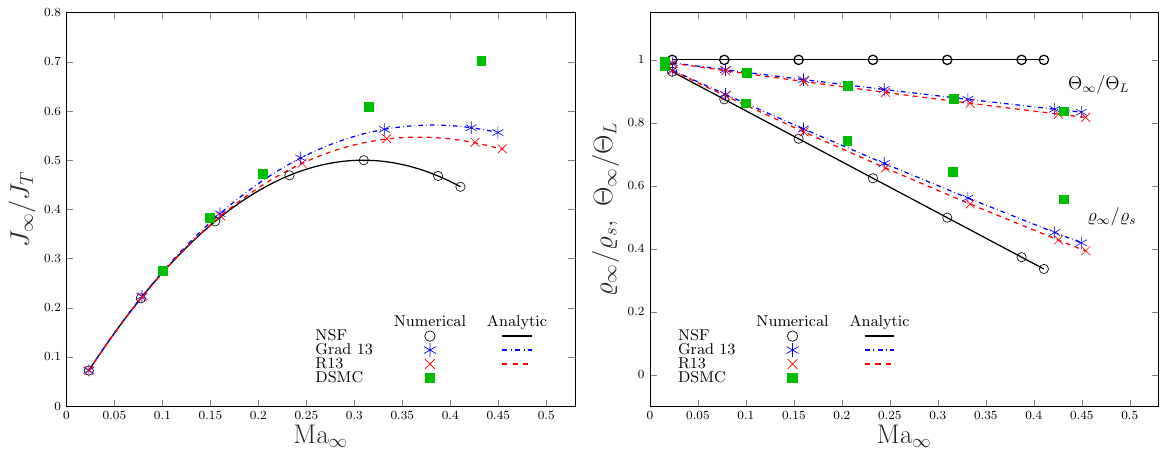}
\caption{\label{fig:flatap1}Left: Far-field mass flux $J_\infty$, normalised with respect to $J_T=\Rho_s\sqrt{\Theta_L/2\pi}$, against Mach number. Right: Far-field density $\Rho_\infty$ and temperature $\theta_\infty$ against Mach number. Comparison between DSMC, NSF, Grad 13 and R13, for the one-dimensional case. Both numerical and analytic results for the moment equations are shown, with numerical results matching analytic results well.}
\end{figure}

Numerical results for the planar case, computed in a 2D domain, match analytic results, which gives us confidence in our computational approach. These are shown in comparison to DSMC results in Figure \ref{fig:flatap1}, showing the far-field mass flux $J_\infty$ normalised with respect to the total emitted (evaporative) mass flux from a given surface, $J_T$, given by \begin{align*}
J_T =\Rho_s \sqrt{\frac{\Theta_L}{2\pi}},
\end{align*}
as well as the ratios $\Rho_\infty/ \Rho_s$ and $\Theta_\infty/ \Theta_L$. The Grad 13 mass flux slightly closer approximates the DSMC results for this case than that of the R13. While we expect R13 to be a better approximation than Grad 13 in the mathematical limit as $\text{Kn}\to 0$, as it is higher-order in Knudsen number, at finite Knudsen numbers there is nothing to prevent Grad 13 fortuitously giving better agreement with the DSMC than R13.  Notably, from an engineering perspective, what we will see throughout the article is that there is a general trend for R13 to be the most accurate method, most notably in Section \ref{sec:knudanal} and Figure \ref{fig:knudsenplots}.

{\parindent0pt
\subsubsection*{Numerical Results in Comparison to DSMC for the General (2D) Case}
}

\begin{figure}
\includegraphics[width=\textwidth,trim={0 0cm 0 0cm}]{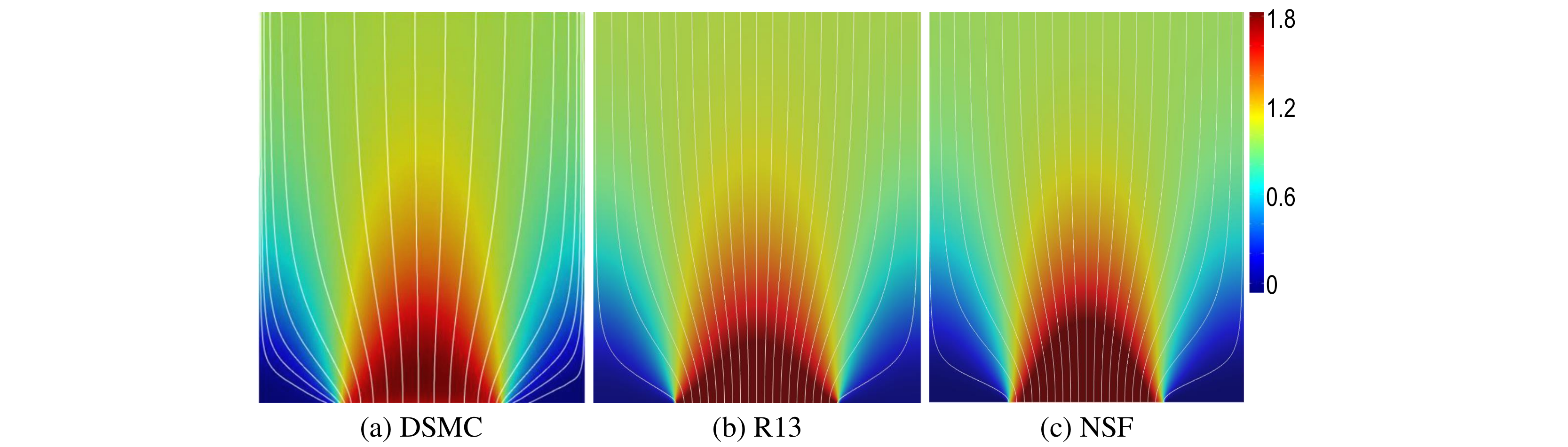}
\caption{\label{fig:vsurfs}Flow-fields near the interface showing the normalized $y$-velocity, $v/v_\infty$, and velocity streamlines for the flat meniscus cases, $\eta = 0$, with porosity $\phi=0.5$, at Kn=0.05. The far-field velocity is $v_\infty=0.2172$, corresponding to $\V_\infty =60$ m/s. Three models are shown; (a) DSMC from \citep{Benzipaper2}, (b) R13, and (c) NSF.}
\end{figure}
\begin{figure}
\includegraphics[width=\textwidth,trim={0 0cm 0 0cm}]{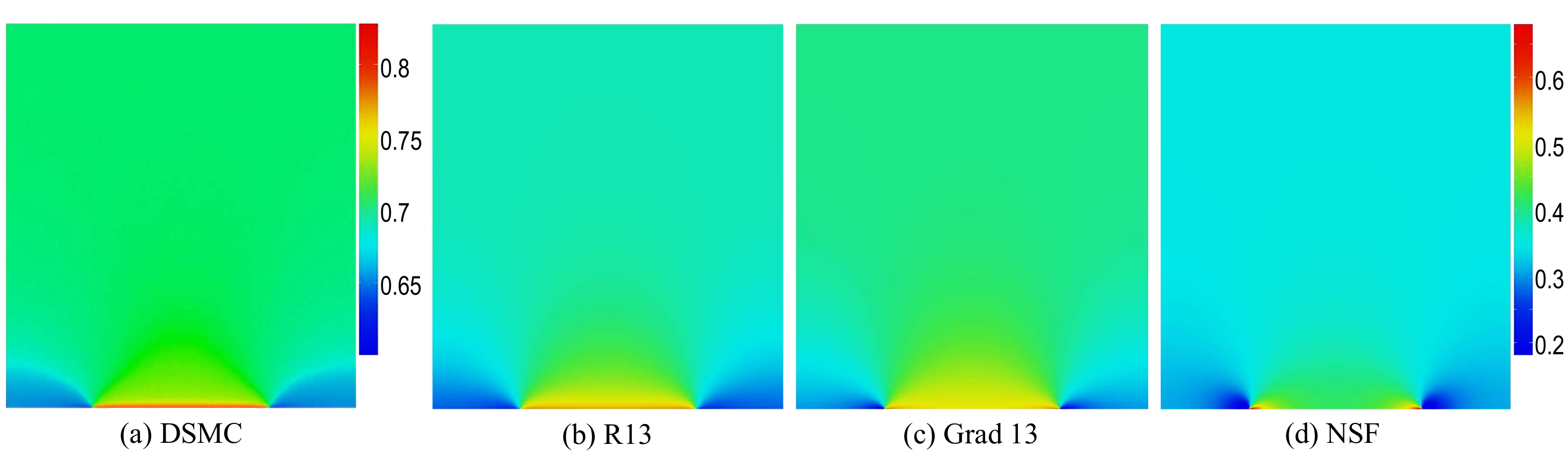}
\caption{\label{fig:rsurfs}Surface plots near the interface showing the density ratio $\Rho/\Rho_s$, for the flat meniscus case, $\eta = 0$, with porosity $\phi=0.5$, at Kn=0.05. The far-field velocity is $v_\infty=0.2172$, corresponding to $\V_\infty =60$ m/s. Four models are shown; (a) DSMC from \citep{Benzipaper}, (b) R13, and (c) Grad 13, and (d) NSF.}
\end{figure}

A plot of the normalised y-component of velocity over a portion of the domain is shown in Figure \ref{fig:vsurfs} for a representative case where both $v_\infty = 0.2172$ and $\kn=0.05$ are relatively low, so we expect the models to agree; for all cases the meniscus is flat ($\eta=0$) and the porosity is a half ($\phi=0.5$). Results for three different models are presented (DSMC from \citep{Benzipaper,Benzipaper2}, R13 and NSF from left to right), overlaid with velocity stream lines.
Grad 13 gives similar results to the R13 model, and is therefore not shown. All moment models reproduce the DSMC velocity field well.

\begin{figure}
\includegraphics[scale=1.289, trim={0 0cm 0 0cm}]{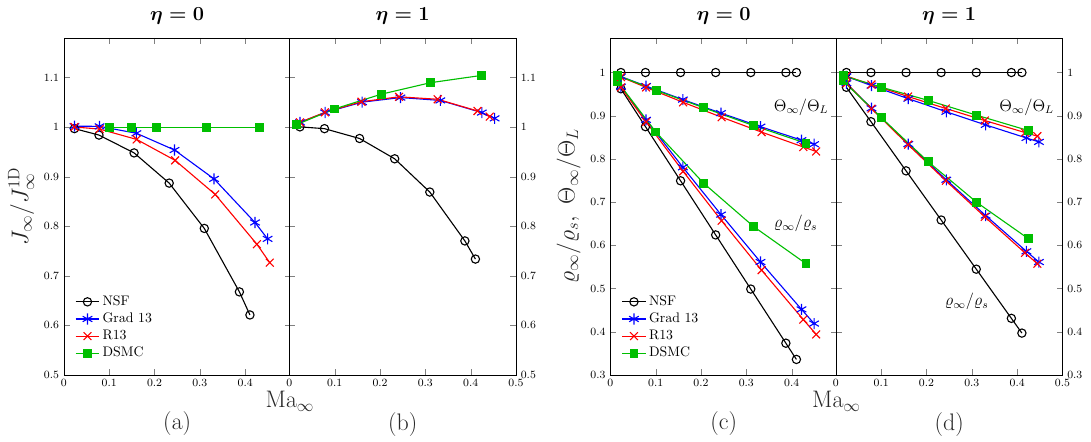}
\caption{\label{fig:ap1}Normalised far-field parameters against far-field Mach number for the full-porosity case ($\phi=1$), at Kn$=0.05$. Shown are $J_\infty / J_\infty^{1\text{D}}$ for (a) the flat meniscus, (b) the semi-circular meniscus; and $\Rho_\infty/ \Rho_s$, $\Theta_\infty/ \Theta_L$ are shown for (c) the flat meniscus, (d) the semi-circular meniscus.  Comparison between DSMC, NSF, Grad 13 and R13.}
\end{figure}
\begin{figure}
\includegraphics[scale=1.289, trim={0 0cm 0 0cm}]{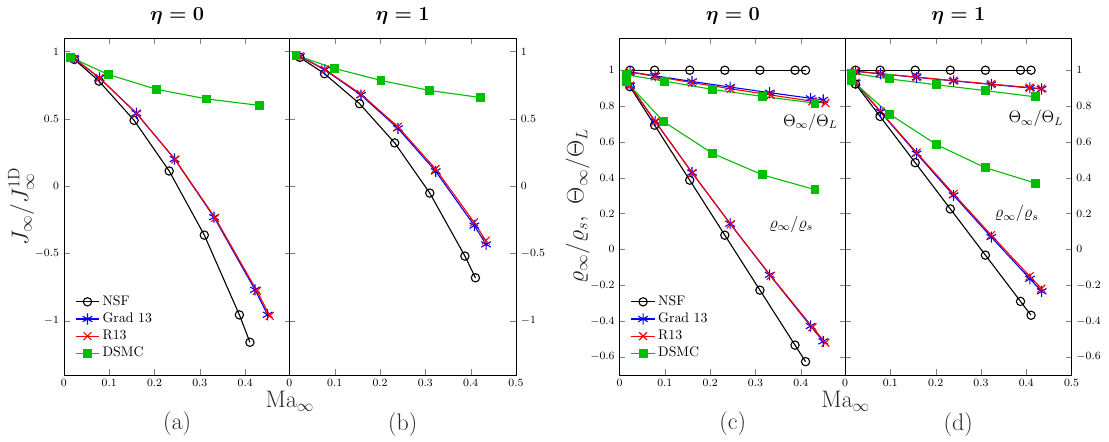}
\caption{\label{fig:ap05}Normalised far-field parameters against far-field Mach number for the half-porosity case ($\phi=0.5$), at Kn$=0.05$. Shown are $J_\infty / J_\infty^{1\text{D}}$ for (a) the flat meniscus, (b) the semi-circular meniscus; and $\Rho_\infty/ \Rho_s$, $\Theta_\infty/ \Theta_L$ are shown for (c) the flat meniscus, (d) the semi-circular meniscus.  Comparison between DSMC, NSF, Grad 13 and R13.}
\end{figure}

The density flow field is shown in Figure \ref{fig:rsurfs}, for all four models, for the same case. 
Grad 13 and R13 show good agreement with DSMC for the variation of density, while NSF densities are clearly inaccurate.
Though the plots do not show it, NSF sees negative values of density around the contact line between the meniscus, wall and vapour. This is caused by a singularity on the wall side of the wall-liquid-vapour contact line, and is discussed in Appendix \ref{sec:singularity}.
The singularity, however, is a local artifact, and does not affect far-field parameters. Promisingly, the higher moment equations, Grad 13 and R13, do not appear to exhibit singular behaviour; see Figure \ref{fig:nsfvG13} in Appendix \ref{sec:singularity} for a comparison of NSF and Grad 13.

For the assessment of far-field mass flux, it is convenient to normalise computed mass flux values with respect to analytic results from the one-dimensional case derived by Labuntsov and Kryukov \cite{1DMF}:
\begin{equation}\label{eqn:J1D}
J_\infty^{\mathrm{1D}} = \vinf \left (2 C_s ^2 + \vinf^2 - \sqrt{4 C_s^2 \vinf^2 + \vinf^4} \right ) \Big / 2C_s^2,
\end{equation}
where $C_s=0.6\sqrt{2\Theta_L}$. We again assess the ratios $\Rho_\infty/ \Rho_s$ and $\Theta_\infty/ \Theta_L$.

Figure \ref{fig:ap1} show the results for the limiting case of $\phi\to 1$, corresponding to a meniscus suspended by plate-like posts, for two cases of curvature, $\eta=0$ and $\eta=1$.
Computed far-field mass flux values for the two 13-moment systems (Grad 13 and R13), shown in Figures \ref{fig:ap1} (a) and (b), are in relatively good agreement with DSMC compared to NSF, particularly at low Mach numbers. 
For $\eta \neq 0$, where a two-dimensional flow is generated (and thus parameters have $x$-dependence) the Knudsen number becomes relevant, and so rarefied-gas dynamics modelling is required, even at relatively low Knudsen numbers. 
For this reason, the 13-moment systems see excellent agreement for the semicircular meniscus, while NSF deviates significantly (for example, showing the wrong qualitative trend in Fig. \ref{fig:ap1} (b)).

Far-field density and temperature both show good agreement with DSMC for the 13-moment systems, and reasonable agreement with DSMC for the NSF system; see Figures \ref{fig:ap1} (c) and (d). 
In particular, the initial slopes, corresponding to $\mathrm{Ma}_\infty \approx 0$, of both densities and the temperatures for both R13 and Grad 13 are in excellent agreement with DSMC. 
As a result, low-Mach number results have near-perfect agreement in both density and temperature. 
The NSF system suffers in this respect, especially for $\eta=1$, since the initial slope in density doesn't agree with the higher-order systems (fig. \ref{fig:ap1} (d)). 
Thus, even at low-Mach numbers we have significant error for the NSF system. 
This in turn means that higher Mach number results are even further away from DSMC results. 

Figure \ref{fig:ap05} shows the results for half porosity, $\phi=0.5$.
We see large discrepancies from DSMC values for computed far-field density, and thus mass flux, for all three moment models. 
The introduction of walls into the simulation produces a strongly two-dimensional flow, particularly near the interface, as seen in Figures \ref{fig:vsurfs} and \ref{fig:rsurfs}. 
This calls for more accuracy in both bulk and boundary condition equations: for the one-dimensional case, only a small amount of terms are removed upon linearisation, while in the two-dimensional case, linearisation removes many more terms. This might explain the discrepancies between the moment equations and DSMC seen for the linear results in Figure \ref{fig:ap05}, as DSMC inherently inlcudes the non-linearities.
Moreover, with linear boundary conditions and bulk equations, the parameters $\rho_\infty$ and $\theta_\infty$ appear to decrease linearly with Mach number.
We saw for the one-dimensional case in the form of the analytic solutions that these parameters are linear in $v_\infty$, which, for small temperatures variations, gives an almost linear relation with Mach number.
By means of a Fourier expansion, it can be shown that for the two-dimensional flat meniscus case far-field parameters are also linear in $v_\infty$, see Appendix \ref{sec:fourier}.
In contrast, the DSMC results for density begin to plateau as the Mach number increases. 
This makes it impossible to retain good agreement for high Mach numbers in the present set up. This is particularly prominent in Figures \ref{fig:ap05} (c) and (d), where we see negative densities at moderate Mach numbers, for all moment models.

The far-field temperature for the two 13-moment systems show good agreement with DSMC values for all cases of porosity and meniscus shape, see (c) and (d) in Figures \ref{fig:ap1} and \ref{fig:ap05}. The lack of a temperature jump in the NSF system means no temperature variation is seen at the far-field, and so accuracy is lost with respect to the DSMC.

For all simulations in Figures \ref{fig:ap1} and \ref{fig:ap05}, Grad 13 and R13 show very similar results, as we would expect at the fairly low Knudsen number of 0.05. 
Much larger differences are seen at higher Knudsen numbers, seen in Section \ref{sec:knudanal}. 
We also see that an increase in either porosity or curvature gives an increase in far-field mass flux, apparently due to the fact that there is increased surface area for evaporation.

\mbox{}

{\parindent0pt
\subsubsection*{Discussion}
}
The aim of this paper is to utilise the moment equations to simulate evaporation from a nanoporous membrane. 
An important region for accuracy for this is that of low Mach number flows at low-to-mid Knudsen numbers, $0.05 < \kn < 1$, as DSMC is generally computationally expensive in this regime, sometimes taking weeks to compute two-dimensional flows \citep{DSMClowmach,DSMCweeks,DSMCdays}. 
Macroscopic equations present an efficient alternative, with our simulations from this section taking less than a minute for a given value of $v_\infty$ on a standard laptop. 
For the linear results presented above, we have accurate results up to $\mathrm{Ma}_\infty \approx 0.1$, but beyond this the results are too inaccurate to be used as a substitute for DSMC. 
We would therefore like to increase the range in which we can use the moment equations and be confident in the results.

%Thus far, the moment equations, and in particular the 13 moment systems, have shown good agreement with DSMC results when no walls are present ($\phi=1$).
%However, when walls were introduced, we saw calculated far-field densities differ from DSMC values by a considerable margin.
Our 2D model is aimed at giving a simplistic, analyzable representation of a nanoporous membrane. 
Since we have considered the process to be adiabatic, energy is removed from the system solely by evaporative mass flux, and so the accurate calculation of far-field densities is critical to understanding the energy-dissipating capabilities of a nanoporous membrane for a given configuration. 
The mass flux-driven energy dissipation would also be seen for the working device \cite{Membranepaper}, and so while our simplified 2D model will not provide exact data corresponding to a cooling device, particularly as a confining 'ceiling' is likely to be present in this case, it will provide general trends which may aid the design of such a device.

To improve the accuracy of the macroscopic approach, we could consider even more moments, since these would theoretically approximate the DSMC model to higher accuracy \citep{Strbook,rationalET,manymoments}. 
While work has been done to understand the number of moments required for accurate simulation of a given problem \citep{koellermeier2019error,abdelmalik2017error}, the computational expense and modelling complexity added for an increased number of moments, particularly for two- and three-dimensional flows, makes this strategy hard to justify \citep{Strbook,momentssummary}. Notably, a promising new approach, based on using a different number of moments in different regions of the domain \cite{torrilhon2017hierarchical}, has the potential to overcome these limitations and should be the focus of future work in this field.

However, any linearised system of higher moment equations will still exhibit linear relations between far-field density and $v_\infty$ for the flat meniscus case: for the one-dimensional case with unitary porosity, we will always be able to obtain a boundary condition system of the form $\mathbf{b} = A \mathbf{U},$
where $A$ has no $v_\infty$-dependence, and the linear, two-dimensional analysis done for arbitrary porosity in Appendix \ref{sec:fourier} works for arbitrary moment systems. 
Thus high Mach number accuracy will be sacrificed for low Mach number accuracy, or vice versa. 
Since the problem seems to stem from our linearity assumption, a first-step improvement is to introduce non-linearity in the boundary conditions, which we now discuss.

\mbox{}

\section{\label{sec:semilin}Extension to non-linear boundary conditions}
Here we argue the case for relaxing the linearity assumption of a component of velocity at the boundary, and then give the modified boundary conditions. Notably, the bulk equations remain linear. We then present the results using these boundary conditions, including a comparison to the linear case.

At the far-field we have a constant velocity of $v_\infty$. When walls are introduced at the interface ($\phi<1$), the continuity equation dictates that there must be some value of $v_2$ such that $v_2>v_\infty$ at the meniscus. 
Therefore, to retain accuracy at higher values of $v_\infty$ (and hence higher Mach numbers) it is prudent to assume that $v_2$ is no longer small along this boundary.

Insertion of (\ref{eqn:linearisationparams}) into (\ref{eqn:evapbc})-(\ref{eqn:highbc3}) with the above assumption, and setting $\chi=1$, gives the following non-linear boundary conditions:

\begin{equation}\label{eqn:nlevapbc}
\tag{\ref*{eqn:evapbc}${}^{\prime\prime}$}
v_{n}(1+\rho)=\frac{\vartheta}{2-\vartheta}\sqrt{\frac{2}{\pi }}\left(p_s-\Pi-\frac{1}{2}\theta_L\right),
\end{equation}
\begin{equation}\label{eqn:nlslipbc}
\tag{\ref*{eqn:slipbc}${}^{\prime\prime}$}
\begin{aligned}
\sigma_{t  n}=&-\sqrt{\frac{2}{\pi }}  \left(v_{t }+\frac{1}{5} q_{t }+\frac{1}{2} m_{t  n n}\right) - v_n v_t,
\end{aligned}
\end{equation}
\begin{equation}\label{eqn:nltempjumpbc}
\tag{\ref*{eqn:tempjumpbc}${}^{\prime\prime}$}
\begin{aligned}
q_{n}=-\sqrt{\frac{2}{\pi }} &\left(2 \left(\theta-\theta_{L}\right)+\frac{1}{2}\sigma_{n n}+\frac{5}{28} R_{n n}\right)-\frac{1}{2}v_n \Big[ \wp_L+5(\theta-\theta_L)\Big],
\end{aligned}
\end{equation}
\begin{equation}\label{eqn:nlhighbc1}
\tag{\ref*{eqn:highbc1}${}^{\prime\prime}$}
\begin{aligned}
m_{n n n}=\sqrt{\frac{2}{\pi }}& \left(\frac{2}{5}\left(\theta-\theta_{L}\right)-\frac{7}{5} \sigma_{n n}-\frac{1}{14} R_{n n}\right)-\frac{2}{5}v_n\wp_L,
\end{aligned}
\end{equation}
\begin{equation}\label{eqn:nlhighbc2}
\tag{\ref*{eqn:highbc2}${}^{\prime\prime}$}
\begin{aligned}
m_{t  t  n}=-\sqrt{\frac{2}{\pi }}& \left( \sigma_{t  t }+\frac{R_{t  t }}{14} +\frac{1}{5} \left(\theta-\theta_{L}\right)-\frac{1}{5} \sigma_{n n}\right)+\frac{1}{5} v_n\wp_L,
\end{aligned}
\end{equation}
\begin{equation}\label{eqn:nlhighbc3}
\tag{\ref*{eqn:highbc3}${}^{\prime\prime}$}
\begin{aligned}
R_{t  n}= \sqrt{\frac{2}{\pi }}&\left(v_t-\frac{11}{5} q_{t }-\frac{1}{2} m_{t  n n}\right)+v_nv_t,
\end{aligned}
\end{equation}
where $\wp_L = 1+\rho+\theta_L$ and $\Pi =\rho +\frac{1}{2}\theta+\frac{1}{2} \sigma_{n n}-\frac{1}{28} R_{n n}.$

The bulk equations remain the same for each of the three moment systems; only the boundary conditions are changed, with \eqref{eqn:linevapbc}-\eqref{eqn:linhighbc3} replaced by \eqref{eqn:nlevapbc}-\eqref{eqn:nlhighbc3}.

{\parindent0pt
\subsubsection*{Analytic Result for a Planar Evaporative Interface (1D Case) with Non-linear Boundary Conditions}
}
We can again solve the system analytically for the case $\phi=1, \eta=0$. The general solution (containing unknown constants of integration) is unchanged from (\ref{eqn:gensol}) from the linear case. 
The boundary system 
\begin{align}\label{eqn:matrixsystem}
\tag{$**$} 
\mathbf{b} = A_{\text{n.l.}} \mathbf{U}
\end{align}
changes from the linear case with the new boundary conditions. The vectors $\mathbf{b}$ and $\mathbf{U}$ are unchanged, however the matrix $A$ becomes
\begin{align*}
A^{[\mathrm{R}13]}_{\text{n.l.}} = \left [\begin{matrix}
-\eip- \vinf & -\frac{1}{\sqrt{2\pi}} & \frac 35 \vinf + \frac{3}{5\sqrt{2\pi}}  \\
-\frac{\vinf}{2} & -\frac 52\vinf - \frac{4}{\sqrt{2\pi}} & \frac{13}{10}\vinf + \frac{3}{5\sqrt{2\pi}} \\
-\frac{2\vinf}{5} & \frac{4}{5\sqrt{2\pi}} & \frac{6}{25}\vinf-\frac{39}{25} \eip -\frac{ \sqrt{30}}{5}
\end{matrix} \right ]
\end{align*}
and $A^{[\mathrm{G}13]}_{\text{n.l.}}$ is the upper left $2\times 2$ submatrix of $A^{[\mathrm{R}13]}_{\text{n.l.}}$, and $A^{[\mathrm{NSF}]}_{\text{n.l.}}$ the upper left entry of $A^{[\mathrm{R}13]}_{\text{n.l.}}$. The non-linear assumption does not introduce a non-constant system, but introduces $v_\infty$-dependence on the right of (\ref{eqn:matrixsystem}), extending the boundary condition system obtained by Struchtrup \citep{StrEvapbc}: the matrix $A^{[\mathrm{R}13]}_{\text{n.l.}}$ can be written as 
\begin{align*}
A^{[\mathrm{R}13]}_{\text{n.l.}} = A^{[\mathrm{R}13]}_{\text{lin.}} + C v_\infty, \qquad  C= \left [\begin{matrix}
-1 & -0 & \frac 35 \\[3pt]
-\frac{1}{2} & -\frac 52 & \frac{13}{10}\\[3pt]
-\frac{2}{5} & 0 & \frac{6}{25}
\end{matrix} \right ]
\end{align*}
where $A^{[\mathrm{R}13]}_{\text{lin.}} $ is the matrix from (\ref{eqn:bcsystem}) in Section \ref{sec:linear}.
This leads to a non-linear relation between the solutions and $v_\infty$, given in Appendix \ref{sec:nonlinan}.
The solutions exhibit asymptotic behaviour, approaching a finite value as $v_\infty\to \infty$, which leads to greatly improved results in the subsonic region we are concerned with, $\text{Ma}_\infty<1$. 
Notably, for these solutions, densities remain positive for all $\mathrm{Ma}_\infty>0$. 
In contrast, analytic results for the linear boundary condition system from Section \ref{sec:linear} see negative far-field densities for all Mach numbers above a certain value ($\mathrm{Ma}_\infty > 0.618$ for NSF, $\mathrm{Ma}_\infty > 0.8357$ for Grad 13, and $\mathrm{Ma}_\infty > 0.8092$ for R13).

Numerical results once again match well with the analytic results, as now described.

\mbox{}

{\parindent0pt
\subsubsection*{Numerical Results in Comparison to DSMC for the General (2D) Case with Non-linear Boundary Conditions}
}

\begin{figure}
\includegraphics[scale=1.289, trim={0 0cm 0 0cm}]{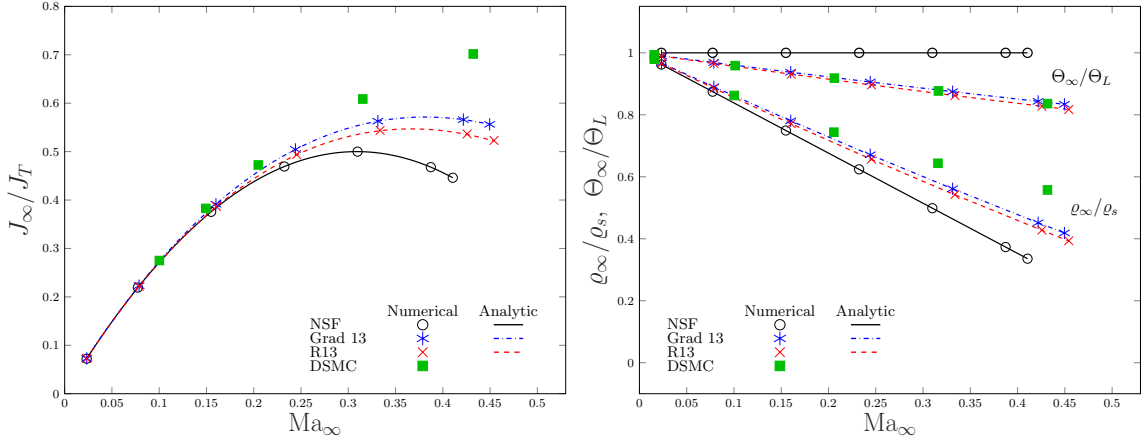}
\caption{\label{fig:ap05lnlcomp}Left: Far-field mass flux $J_\infty$, normalised with respect to $J_\infty^{1\text{D}}$, against Mach number. Right: Far-field density $\Rho_\infty$ as a ratio of saturation density, against Mach number. Comparison between DSMC, and the linear and non-linear boundary conditions models of the two 13-moment system, for the half-porosity case ($\phi=0.5$), at Kn$=0.05$.}
\end{figure}

We assess all parameters in the same way as for the linear case, with mass flux normalised by $J_\infty^{1D}$, given by (\ref{eqn:J1D}). 
For a direct comparison with the linear case, we look at the density profiles for the case of half porosity and a flat meniscus, $\phi=0.5, \eta=0$. 
Figure \ref{fig:ap05lnlcomp} shows this result. 

We can see clearly the improvement of the new model and thus the importance of the boundary conditions to the system's behaviour, as we may expect at Kn=0.05 where bulk non-equilibrium effects are less pronounced. 
We see the effect non-linearity has on the relationship between Mach number and density, with asymptotic behaviour similar to that discussed in the one-dimensional case seen as Mach number increases. 
Moreover, the initial slope of the density plot for $\mathrm{Ma}_\infty \approx 0$ is in very good agreement with that of DMSC, meaning, along with better behaviour for larger Mach numbers, we see even better agreement at lower Mach numbers than with the linear case. 

\begin{figure}
\includegraphics[scale=1.289, trim={0 0cm 0 0cm}]{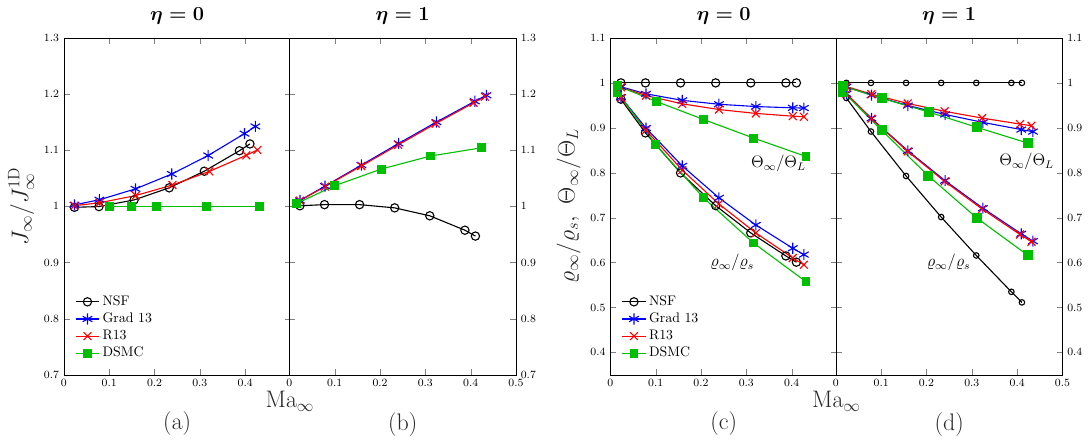}
\caption{\label{fig:ap1nl}Normalised far-field parameters against far-field Mach number for the full-porosity case ($\phi=1$) and Kn$=0.05$. Shown are $J_\infty / J_\infty^{1\text{D}}$ for (a) the flat meniscus, (b) the semi-circular meniscus; and $\Rho_\infty/ \Rho_s$, $\Theta_\infty/ \Theta_L$ for (c) the flat meniscus, (d) the semi-circular meniscus. Comparison between DSMC, and non-linear boundary condition models for NSF, Grad 13 and R13.}
\end{figure}

\begin{figure}
\includegraphics[scale=1.289, trim={0 0cm 0 0cm}]{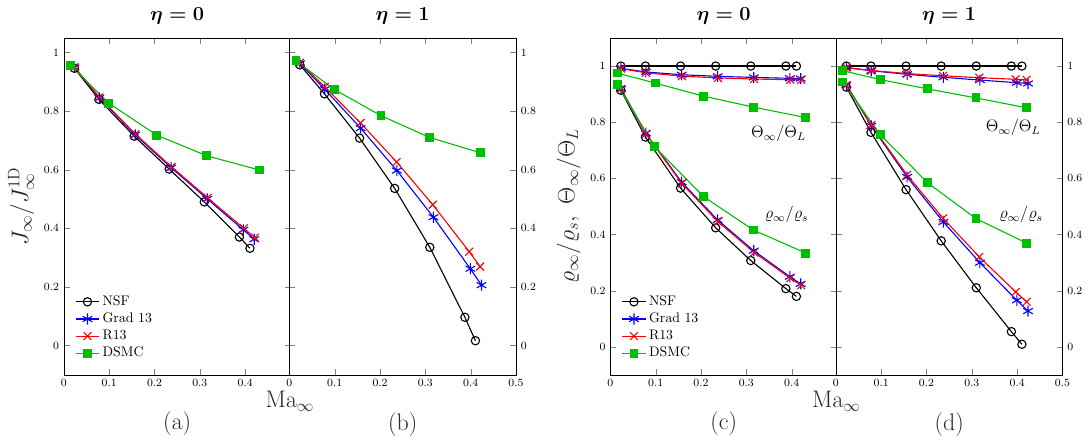}
\caption{\label{fig:ap05nl}Normalised far-field parameters against far-field Mach number for the half-porosity case ($\phi=0.5$), at Kn$=0.05$. Shown are $J_\infty / J_\infty^{1\text{D}}$ for (a) the flat meniscus, (b) the semi-circular meniscus; and $\Rho_\infty/ \Rho_s$, $\Theta_\infty/ \Theta_L$ for (c) the flat meniscus, (d) the semi-circular meniscus. Comparison between DSMC, and non-linear boundary condition models for NSF, Grad 13 and R13.}
\end{figure}

Further simulation data for combinations of $\phi=0.5, 1$ and $\eta=0,1$, can be seen in Figures \ref{fig:ap1nl} and \ref{fig:ap05nl}. 
The improvement in accuracy allows us to look at Mach numbers up to $\text{Ma}_\infty \approx 0.45$, and still retain reasonable accuracy. 
In contrast, the fully linear case saw negative far-field densities at $\text{Ma}_\infty \approx 0.25$.
Figures \ref{fig:ap1nl} (a) and (c) show the results from the one-dimensional case where analytic results were obtained. Analytic results are not shown, but match well with the numerical model, which was computed in a two-dimensional domain.
The results give good agreement with the DSMC results for all moment systems. The non-linearity of the solution allows calculated values to closely follow the DSMC values for higher Mach numbers, compared to the linear one-dimensional case, Figure \ref{fig:ap1} (c), where density started with good agreement, but fell away from DSMC values as the Mach number increased.

For the linear case, when walls were introduced, we saw far-field density results diverge greatly from DSMC values at higher Mach numbers. The non-linear boundary conditions give greatly improved results here over the linear case, for both cases of curvature, seen in Figure \ref{fig:ap05nl} (compared to the linear boundary condition case, Figure \ref{fig:ap05}).
We also see that for all results, when $\mathrm{Ma}_\infty \approx 0$ the slopes of density are very accurate compared to the DSMC plots, yielding excellent results for small Mach numbers. 
As discussed in the linear case, low Mach number flows are an important region of accuracy for the higher order moment systems.

The results presented give good results up to modest Mach numbers for all cases of porosity and meniscus shape considered.
This allows us to greatly extend the range of Mach numbers in which we can have confidence in the moment equations results compared to the linear case, where we could only be confident up to $\mathrm{Ma}_\infty \approx 0.1$. 
For non-linear boundary conditions, the 13-moment systems give accurate results up to $\mathrm{Ma}_\infty \approx 0.3$. Therefore, at least for $\kn=0.05$, the 13-moment systems make a good, efficient substitute for DSMC to reduce computational expense on simulating nanoporous membrane dynamics.

\mbox{}

\section{\label{sec:knudanal}Knudsen-Number Analysis}

\begin{figure}
\includegraphics[scale=1.45, trim={0 0cm 0 0cm}]{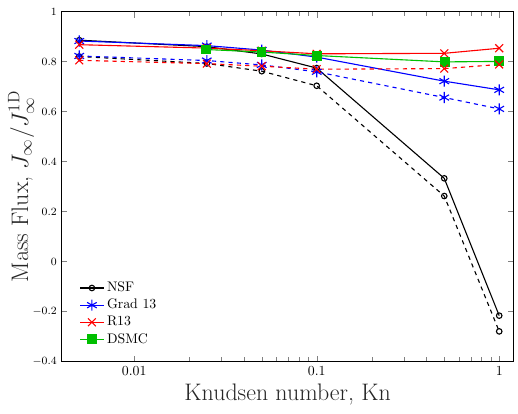}
\caption{\label{fig:knudsenplots}Far-field mass flux $J_\infty $ normalised with respect to $J_\infty^{1D}$, against Knudsen number $\kn$. 
Comparison between DSMC, NSF, Grad 13 and R13, for the half porosity ($\phi=0.5$), flat meniscus ($\eta=0$) case, with $\V_\infty=30$m/s. Solid lines represent non-linear boundary conditions models, dashed lines represent linear models.}
\end{figure}

So far, we have focussed on $\kn=0.05$. 
The moment equations are approximations of the Boltzmann equation with increasing accuracy in Knudsen number, with NSF being first order accuracy, Grad 13 second order, and R13 third order \citep{Strbook,Strknud}. For the flat meniscus case, in the limiting case of infinitely thin walls, $\phi\to 1$, the Knudsen number is irrelevant when evaluating far-field parameters, and so we expect these models to be similar, with differences coming from the different boundary condition models for each set of equations. When either $\eta>0$ or $\phi<1$, the solutions become dependent on the Knudsen number. 
This is seen in Figure \ref{fig:knudsenplots}, which gives the results of the normalised far-field mass flux for increasing Knudsen number for both linear and non-linear boundary condition models, for the representative flat meniscus, half porosity case, and a far-field velocity of $\V_\infty=30$m/s, corresponding\footnote{Assuming a molar mass of $M=0.02896 $kg/mol and temperature of $273$ K.} to $v_\infty = 0.1072$.
All macroscopic models retain accuracy up to around Kn=0.1. For Kn$>0.1$, the NSF solutions diverge greatly from DSMC.
Meanwhile, the Grad and R13 equations, which are of higher-order accuracy in the Knudsen number, maintain reasonable accuracy to the DSMC values.
We do see the Grad 13 far-field mass flux gradually decreasing, dropping from $J_\infty / J_\infty^{1D} = 0.882$ at $\kn=0.005$, to $J_\infty / J_\infty^{1D} = 0.687$ at $\kn=1$ for the non-linear case. 
On the other hand, R13 maintains a value of $J_\infty / J_\infty^{1D}$ between 0.868 and 0.831 for all Knudsen numbers for the non-linear boundary conditions case, hugging the DSMC values between 0.848 and 0.801.
We also see that as $\kn\to 0$, all moment solutions converge, as expected.

\begin{figure}
\includegraphics[scale=1.289, trim={0 0cm 0 0cm}]{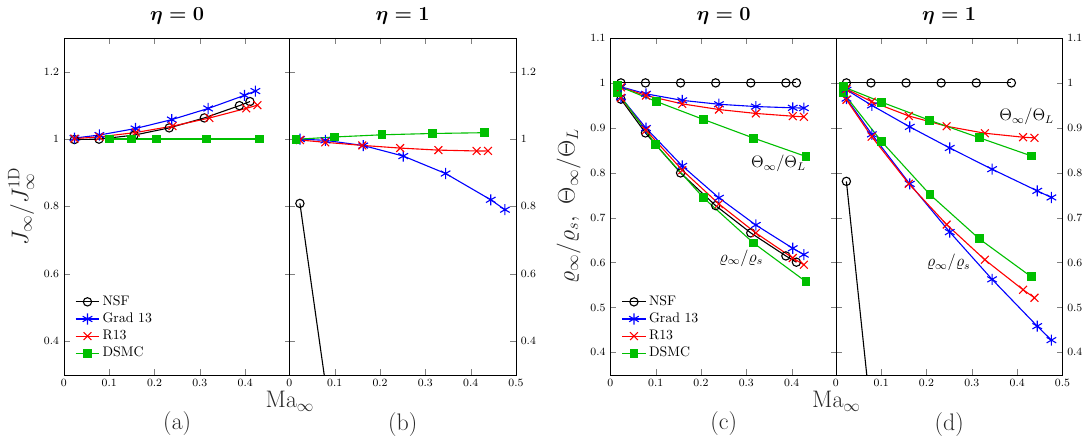}
\caption{\label{fig:ap1kn1}Normalised far-field parameters against far-field Mach number for the full-porosity case ($\phi=1$), at Kn$=1$. Shown are $J_\infty / J_\infty^{1\text{D}}$ for (a) $\eta=0$, (b) $\eta=1$; and $\Rho_\infty/ \Rho_s$, $\Theta_\infty/ \Theta_L$ are shown for (c) $\eta=0$, (d) $\eta=1$.  Comparison between DSMC, and non-linear boundary condition models for NSF, Grad 13 and R13. Very large negative densities and mass flux values are seen for NSF, and most of the computed results are not shown.}
\end{figure}

\begin{figure}
\includegraphics[scale=1.289, trim={0 0cm 0 0cm}]{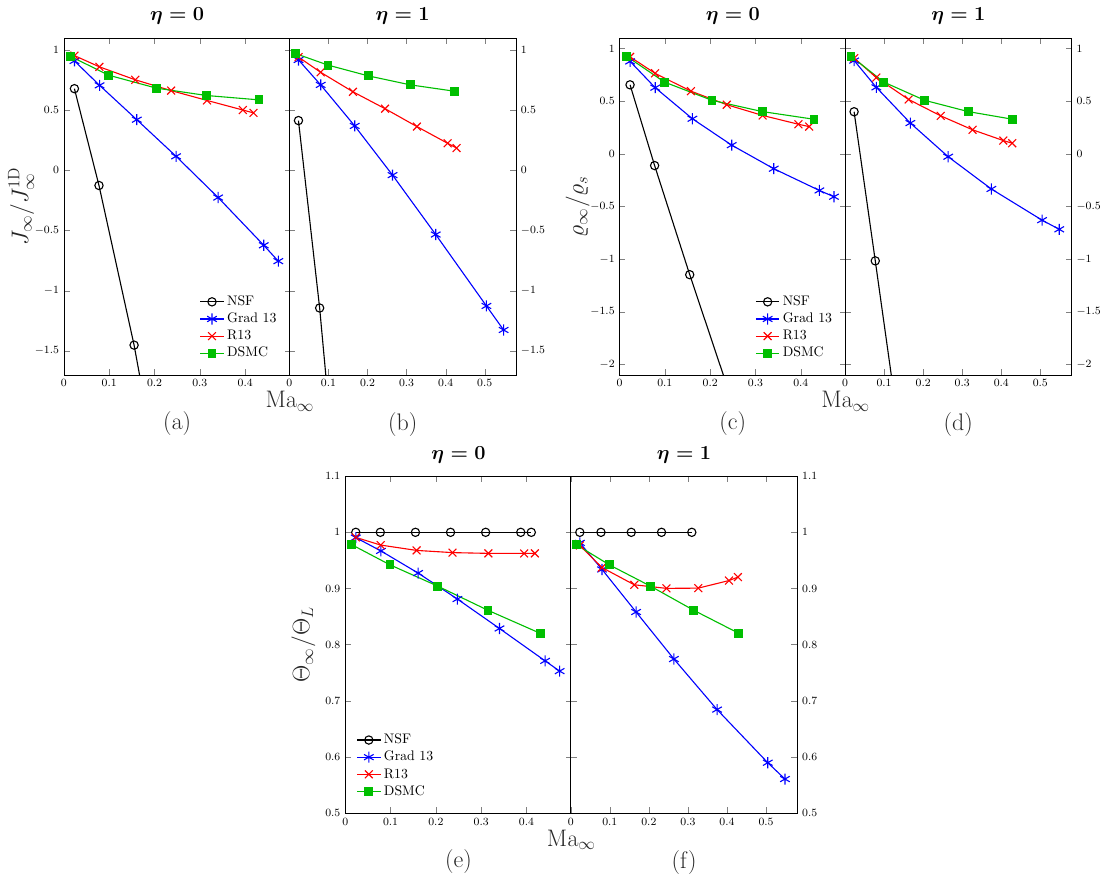}
\caption{\label{fig:ap05kn1}Normalised far-field parameters against far-field Mach number for the half-porosity case ($\phi=0.5$), at Kn$=0.05$. From top to bottom are $J_\infty / J_\infty^{1\text{D}}$, $\Rho_\infty/ \Rho_s$  and $\Theta_\infty/ \Theta_L$ for both cases of meniscus shape.  Comparison between DSMC, and non-linear boundary condition models for NSF, Grad 13 and R13. Very large negative densities and mass flux values are seen for NSF, and most of the computed results are not shown.}
\end{figure}

All far-field parameters are assessed again with respect to Mach number for various membrane configurations, this time for $\kn=1$, in the same way as was done for $\kn=0.05$: mass flux $J_\infty$ is normalised with respect to (\ref{eqn:J1D}), and we assess the density and temperature ratios $\Rho_\infty/ \Rho_s$ and $\Theta_\infty/ \Theta_L$. Simulations are carried out using the non-linear boundary-condition model. Results are shown in Figures \ref{fig:ap1kn1} and \ref{fig:ap05kn1}.


As expected, Figures \ref{fig:ap1kn1} (a) and (c) show no change to the $\kn=0.05$ case, as Knudsen number is irrelevant for far-field parameter computation for the case of a flat meniscus and no walls.
For the semi-circular meniscus case with no walls, shown in Figures \ref{fig:ap1kn1} (b) and (d), the R13 system is still showing good results at $\kn=1$, with all three parameters showing good accuracy in comparison to DSMC. 
Grad 13 is also reasonably accurate in comparison to DSMC, but less so than the R13. 
This is in-line with what we would expect to see at higher Knudsen numbers, since R13 is of higher order accuracy in Knudsen number than Grad 13. 
The NSF system, which is even lower in accuracy in Knudsen number than Grad 13, is completely inaccurate in this regime. The computed density and mass flux values are seen to deviate by large factors from the DSMC results, even for small Mach numbers. 

The case $\phi=0.5$ paints a similar picture; the higher Knudsen number has a large impact on the free stream calculation of density and mass flux, with very large negative values seen in Figures \ref{fig:ap05kn1} (a), (b), (c) and (d) for the NSF system. 
For free-stream velocities $v_\infty > 0.4$, the FEM solver for NSF failed to converge altogether. The reason for this is unclear, but a failure to capture the flow near the singularity discussed in Appendix \ref{sec:singularity} appears to be the most likely cause.
The 13-moment systems again fair much better at the higher Knudsen number than NSF, however Grad 13 sees negative densities at $\text{Ma}_\infty \approx 0.25$ in Figure \ref{fig:ap05kn1} (c), and even sooner in Figure \ref{fig:ap05kn1} (d). 
The R13 system once again is better than the other moment systems, retaining good accuracy up to reasonable Mach numbers.

Changes in the porosity, $\phi$, seem to highlight the difference between the Grad 13 and R13 systems: when $\phi=1$, far-field densities of the two 13-moment systems are comparable, as seen in Figure \ref{fig:ap1kn1} (d); when $\phi=0.5$, however, there are large differences in calculated densities between Grad 13 and R13, as seen in Figures \ref{fig:ap05kn1} (c) and  (d). 
This could be due to the fact that Knudsen layers have a large effect when walls are introduced, as parameters see large variations in values, particularly near the interface; see the flow-field plots in Figure \ref{fig:vsurfs}. An increased number of moments must be considered when dealing with sharp gradients \citep{manymoments,rationalET,torrilhon2017hierarchical}.

\begin{figure}
\includegraphics[width=\textwidth, trim={0 0cm 0 0cm}]{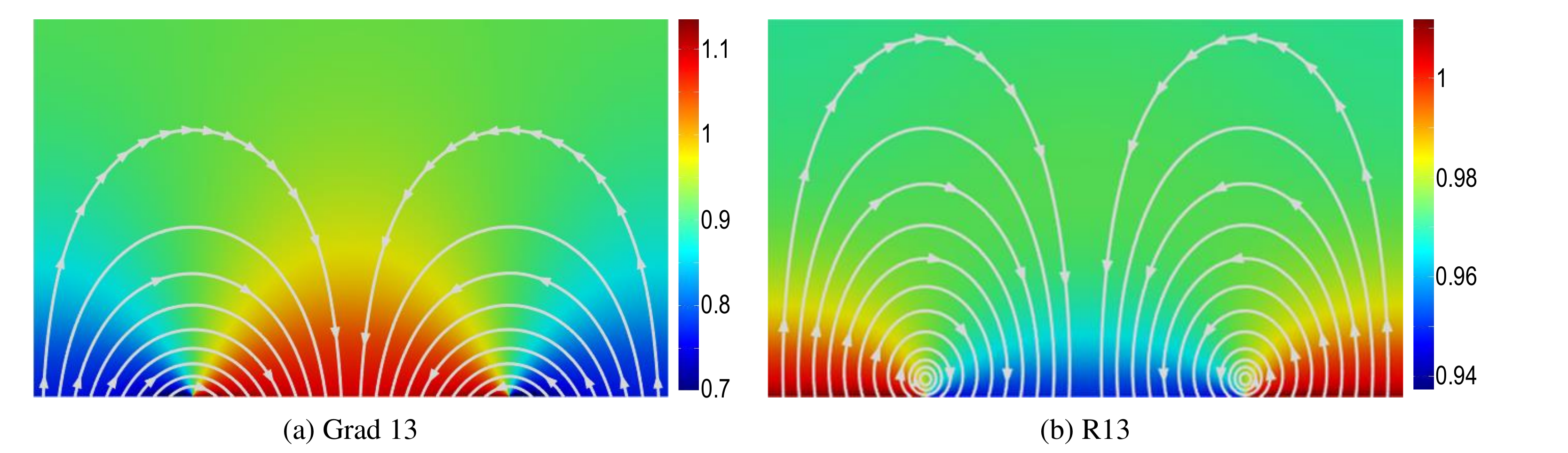}
\caption{\label{fig:fluxflow}Temperature plots near the meniscus for (a) the Grad 13 systems, and (b) the R13 system, overlaid with heat flux streamlines, for the flat meniscus case, with $\phi=0.5$ and $\kn=1$.}
\end{figure}

We can see some of the differences in the two 13-moment systems with the non-linear boundary conditions by looking at the behaviour of temperature and heat flux in the domain. The temperature surface plot, overlaid with heat flux streamlines, is shown in Figure \ref{fig:fluxflow}. 
With $\kn=1$, Figure \ref{fig:fluxflow} (a) shows that the Grad 13 system has a large temperature drop adjecent to the wall, while R13 exhibits larger temperature drops next to the meniscus, Figure \ref{fig:fluxflow} (b). 
The latter is in line with DSMC results, which show larger temperature drops at the meniscus for all Knudsen numbers; see \citep{Benzipaper}, page 14. 
Both moment systems have flux lines emanating from the wall and absorbed at the meniscus. Therefore, while streamlines of heat flux (in general) are directed from hot to cold for the R13 system, we see an inverse Fourier-law prediction by the Grad 13 system, with streamlines of heat flux directed from cold to hot. 

\begin{figure}
\includegraphics[scale=1.45, trim={0 0cm 0 0cm}]{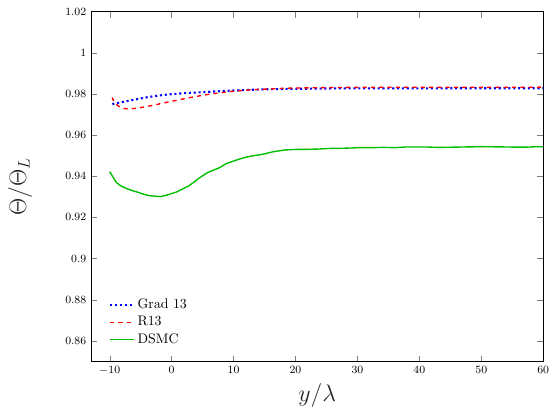}
\caption{\label{fig:yparams}Temperature variation along the extent of of the domain, for the semi-circular meniscus, at $\phi=0.5$ and $\kn=0.05$.}
\end{figure}

Another difference in the two 13-moment systems is the presence of Knudsen layers in the R13 system, which are absent in the Grad 13 system. 
The existence of Knudsen layers is seen in Figure \ref{fig:yparams}, where the temperature ratio $\Theta/\Theta_L$ is plotted along the central vertical axis, $x=0$, for $\kn=0.05$, $\eta=1$ and $\phi=0.5$. 
The Grad 13 profile for temperature is mostly monotonically increasing, especially near the interface. 
Meanwhile, both R13 and DSMC see a decrease in temperature near the interface, and then, after reaching a minimum, we see the monotonically increasing behaviour, where the calculated values plateau to equilibrium. 
The temperature minimum for R13 is attained at $y=-0.35$, corresponding to roughly three mean-free-path lengths from the meniscus. 
Knudsen layers typically extend a few mean-free-path lengths from the boundary, and we also saw that the one-dimensional case had exponential decays on the order of the Knudsen number for R13, and so the drop in temperature can be attributed to Knudsen boundary layer effects. 

Heat flux in the R13 system is also subject to Knudsen layer effects, similar to those seen in the temperature just discussed. Heat flux profiles along the lines $x=\pm \frac 12$ see an extremum at 0.5 mean-free-path lengths and 0.8 mean-free-path lengths from the interface for the tangential and normal components of heat flux respectively.
This results in a sign change inside the domain, leading to the formation of heat flux vortices, seen in Figure \ref{fig:fluxflow} (b). 
These vortices are not seen for Grad 13. Similar behaviour is discussed in the problem of flow past a sphere, where Torrilhon shows the existence of heat flux vortices for the R13 model, which are attributed to the Knudsen boundary layers \citep{R13sphere}. Moreover, Torrilhon showed that such vortices do not exist for a hybrid Stokes-R13 flow, consisting of Stokes bulk equations with R13 boundary conditions.

We have seen the R13 system able to produce accurate results up to Knudsen numbers 1, while Grad 13 gives poor results in this regime, and NSF is completely invalid. 
We saw evidence of Knudsen boundary layers from the R13 system, which are also in DSMC results, while Grad 13 saw no such behaviour, which rationalises R13's superiority, as rarefaction effects play a larger role in the overall flow when the mean-free-path length of particles becomes comparable to the flow's characteristic length scale. 
We can have confidence in the results from R13 with non-linear boundary conditions to be accurate for all  $0<\mathrm{Ma}_\infty <0.3$ and $0\leq \kn \leq 1$. This gives us a large and useful range of membrane configurations which can be simulated accurately and efficiently.

\section{Conclusion}\label{sec:conc}
In this work, the linearised moment equations were used to describe evaporation from a nanoporous membrane. Initially, a fully linear system was considered, in particular with linear boundary conditions, however this failed to give accurate results for Mach numbers above $\text{Ma}_\infty \approx 0.1$.
When retaining non-linearity for velocity in the boundary conditions, we derived an efficient method capable of reproducing DSMC simulation results, effective up to reasonable Mach numbers and Knudsen numbers, which see huge improvements over Navier-Stokes-Fourier results.
The non-linear boundary condition model for the R13 equations is able to produce good results for Mach numbers up to 0.3 and up to $\kn=1$. The NSF is completely unviable in this regime. 
% Moreover, the higher moment equations appear to be singularity free at the wall-liquid-vapour contact point, while the NSF equations appear to contain a singularity at this point.

With the interfacial evaporation processes validated for the moment equations versus DSMC simulations, future work can be devoted to extending the geometry to a more representative version of a nanoporous membrane. 
This could include the introduction of the liquid and solid elements of the membrane using appropriate macroscopic equations, something not possible with DSMC simulations due to the large variations in mean free path length of the different materials.
Introduction of new materials to the simulation requires proper treatment of conservation laws at the interface, including mass, momentum and energy balance, the basics of which can be found in \cite{StrEvapbc} and \citep{Bondbalance}. 
For a full description of the interface, particularly for curved menisci, the surface tension must be introduced, relating the stresses to the meniscus shape \citep{Youngdrop,Thermodynamics}.

In our model we assumed the existence of a far-field where parameters are given space to decay.
For a cooling device designed for micro-scale devices, a more compact geometry is likely, in which a ceiling is brought in, and the gas escapes from the sides due to a cross flow. 
Another direction this project could be taken is to attempt to model this situation. 
The ceiling would complicate the model since the evaporation would no longer be driven by the far-field velocity, and may need to be driven by pressure gradients across the interface. 
Introducing the full geometry with liquid and solid below the interface discussed above could be introduced for the calculation of saturation pressure, since the interplay between saturation pressure and prescribed pressure gradients would play a large role in the evaporation from the meniscus.

Another area of interest would be using the moment equations to simulate the full three-dimensional geometry of a nanopore, something not done yet using DSMC. The efficiency of the macroscopic approach would be very important for this, since large parameter spaces in three dimensions is extremely computationally expensive for DSMC. The two-dimensional results given here provide justification and validation for the use of the 13-moment systems in three-dimensions.

Future research may focus on analysis of the behaviour in pressure/density at the wall-liquid-vapour contact line for the higher moment equations, as they do not appear to exhibit singularities, while the NSF equations do (see Figure \ref{fig:nsfvG13} in Appendix \ref{sec:singularity}). Singularity-free equations are desirable from a physical viewpoint as well as for numerical methods, since mesh refinement can drastically alter local parameter values when a singularity is present.
This study could involve a polar coordinate analysis similar to that performed by Taylor \cite{taylorpaint}, Moffat \cite{moffateddies}, and Nitsche and Parthasarthi \cite{nitscheseep}.

More generally, our findings motivate the hybrid non-linear-boundary linear-bulk equation system which reduces computational complexity from the fully-non-linear approach and yet significantly extends the window of accuracy of the moment equations to higher Mach numbers. Such a hybrid allows application of the method of fundamental solutions, a highly-efficient numerical scheme that employs fundamental solutions to the linear bulk equations (Lockerby  and Collyer \cite{lockerby2016fundamental}, Claydon \emph{et al.} \cite{claydon2017fundamental}). This approach could be useful for a variety of vapour/gas flows where moment equations are considered useful.

\appendix
\section{\label{sec:momderiv}Derivation of the Moment Equations}
The equations are derived from basic considerations of particles in a gas in the 6-dimensional phase space of positions and velocities representing the system. Notation follows that of \cite{Strbook}. We consider the gas to be monatomic, with each particle having mass $m$. The phase space consists of 3 positional directions, $x_i$, and 3 velocity directions, $c_i$. The distribution function $f(x_i, c_i, t)$ is defined such that $f(x_i, c_i, t)\, d \mathbf{x}d \mathbf{c} $ gives the number of particles in the cell $d\mathbf{x} d \mathbf{c}$ at time $t$. One considers the evolution of the distribution function in a given subset $\Omega \subset \mathbb{R}^3$, from which one obtains the Boltzmann equation,
\begin{equation}\label{eqn:boltzmann}
\frac{\partial f}{\partial t}+c_{k} \frac{\partial f}{\partial x_{k}}+G_{k} \frac{\partial f}{\partial c_{k}}=\mathcal{S},
\end{equation}
where $G_i$ are external forces, and $\mathcal{S}=-\frac{\partial }{\partial c_k}(W_k f)$ is the particle interaction term, $W_k$ denoting the intermolecular forces.

 Appropriate integration of $f$ over the velocity space gives the moments. For example, multiplication of $f$ by $m$, the particle mass, and subsequent integration gives the density
\begin{equation*}
\Rho = m\int f\, d\mathbf{c}.
\end{equation*}
Multiplication of $f$ by $c_i$ and subsequent integration gives the momentum
\begin{equation*}
\Rho\V_i = m\int c_i f\, d\mathbf{c},
\end{equation*}

The moments are generated by multiplication of $f$ by polynomials in $c_i$. 
Following Struchtrup \citep{Strbook}, we use the vector of polynomials \[\Psi^{[13]} = m\left \lbrace 1, c_i, \frac 13 C^2, C_{\langle i} C_{j \rangle}, \frac 12 C^2 C_i\right \rbrace \]
to generate our 13 moments. 
Indices in angular brackets denote the trace-free-symmetric part of a tensor \citep{Strbook}, and $C_i = c_i - \V_i$ is the peculiar velocity. 
This gives the 13 moments\footnote{The stress tensor is trace-free and symmetric and so has 5 independent components.} as
\[\Phi^{[13]}= \Rho\left \lbrace 1, \mathcal{V}_i, \Theta, \Sigma_{ij}, \mathcal{Q}_i \right \rbrace.
\]
In the above, $\Theta$ is the temperature in specific energy units, i.e. $\Theta = \frac km T$, where $T$ is the temperature, and $k$ is the Boltzmann constant. 
Additionally, $\Sigma_{ij}$ is the stress tensor, and $\Q_i$ is the heat flux tensor. 
We also study a five-moment system, the NSF system, consisting of the moments $\Phi^{[5]} = \Rho \set{1,\V_i,\Theta}$.

Next, we derive the equations describing the moments $\Phi^{[13]}$. 
We perform a similar process to the one carried out to obtain the moments $\Phi^{[13]}$, but carried out on the Boltzmann equation (\ref{eqn:boltzmann}), rather than the distribution function $f$. 
That is, we multiply (\ref{eqn:boltzmann}) by a polynomial in $c_i$, and then integrate over the velocity space. For instance, multiplication of (\ref{eqn:boltzmann}) by $m$ and subsequent integrating gives the conservation of mass equation
\begin{equation*}
D_t \Rho +\Rho \partial_k \mathcal{V}_k=0.
\end{equation*}

Performing this process for all polynomials in the vector $\Psi^{[13]}$ gives the equations (\ref{eqn:conservationlaws})-(\ref{eqn:heatflux}).

\section{\label{sec:meshpic}Domain finite element mesh}
\begin{figure}
\includegraphics[width=.4\textwidth, trim={0 0cm 0 0cm}]{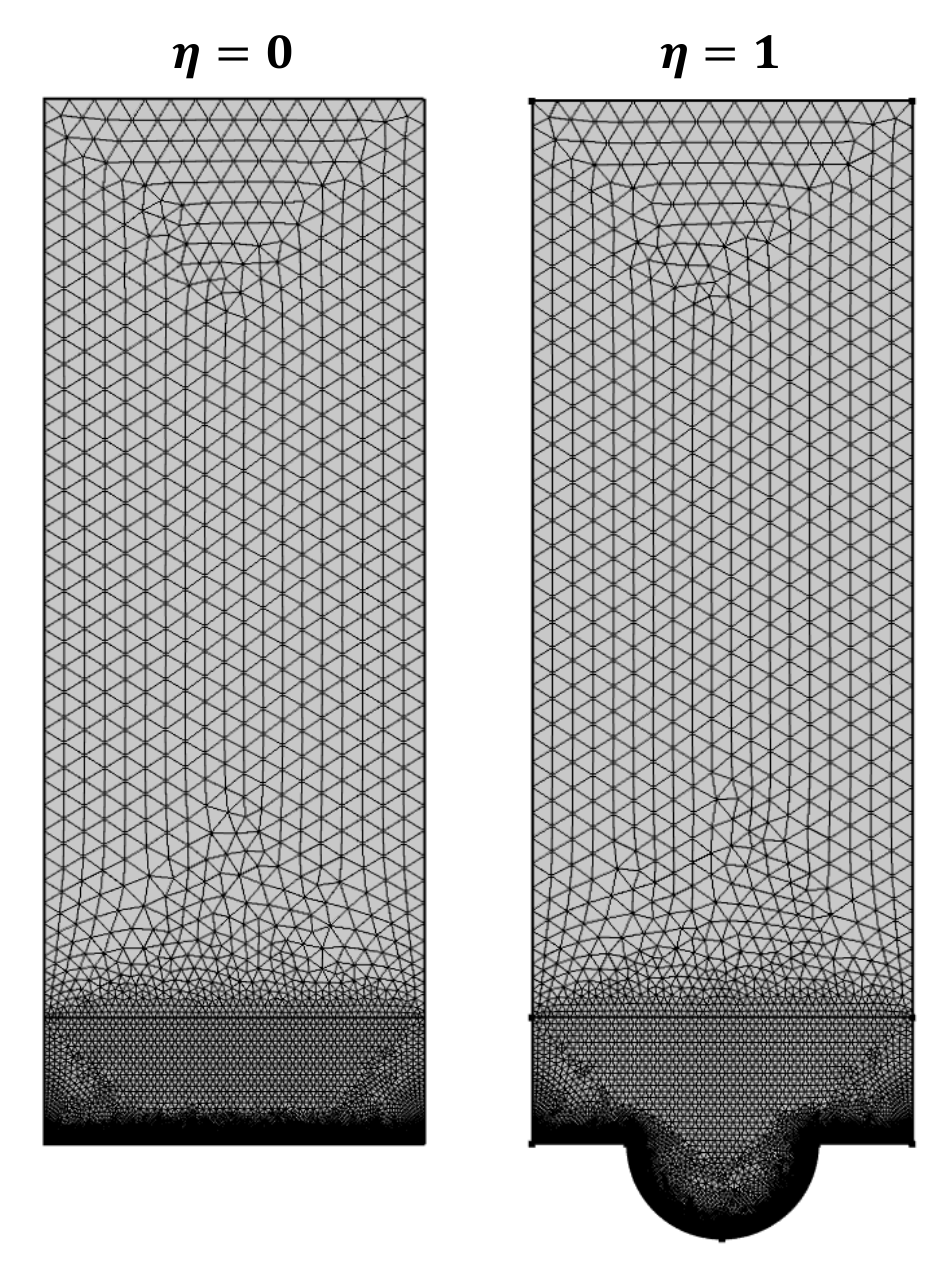}
\caption{\label{fig:meshes}Domain mesh for nanoporous evaporation.}
\end{figure}

The domain mesh for a representative case of $\kn=0.05$ and with half porosity ($\phi=0.5$) is shown in Figure \ref{fig:meshes}, for both the flat meniscus and semicircular meniscus case. The  mesh  consists  of  12369 triangular domain elements and 580 boundary elements for the flat meniscus case, and 14115 triangular domain elements and  637  boundary  elements  for  the  semicircular  meniscus case. The mesh is made finest near the interface where most two-dimensional-type flow occurs. Near the far-field, where the flow is essentially one-dimensional, only a coarse mesh is required

\section{\label{sec:fourier}Fourier decomposition for the two-dimensional case}
Here we examine the structure of the analytic solutions for the flat meniscus with arbitrary porosity. For porosity $\phi=1$, we saw a linear relationship between density and far-field velocity. We will see that this is the case also for arbitrary $\phi$.
We assume parameters take the form 
\begin{align*}
\psi ( x,y) = \sum_{k\in\mathbb{Z}} \psi_k(y) e^{-ik \pi \frac{x}{\ell}},
\end{align*}
where $\ell = \frac 12 L + W$ is the (half) length of periodicity (the half coming from the factor of 2 in front of $\pi$).

With this, the linearised moment equations (\ref{eqn:linconservationlaws})-(\ref{eqn:linstress}) become an infinite set of ordinary differential systems, one for each mode $k$. For each mode, the (arbitrary $N$) moment system can be reduced to a matrix differential system of the form
\begin{equation}\label{eqn:AppODE}
\frac{d}{dy}\varphi^{[N]}_k = Q_k \varphi^{[N]}_k,  \qquad \text{(no sum over } k)
\end{equation}
for the vector of $N$ moments $$\varphi^{[N]}_k = \left \lbrace \rho_k(y), v_{i,k}(y), \theta_k(y), \sigma_{ij,k}(y), q_{i,k}(y),... \right \rbrace^t.$$ Here, $Q_k$ is a matrix dependent on $k$, $\ell$ and $\kn$.

The general solution to \eqref{eqn:AppODE} has the form 
\begin{align*}
\varphi^{[N]}_k  = \exp (Q_k y) \mathbf{C}'_k = P_k \exp (J_k y) \mathbf{C}_k
\end{align*}
where $J_k$ is the Jordan matrix of $Q_k$, $P_k$ is the matrix of generalised eigenvalues, and $\mathbf{C}_k = P_k^{-1} \mathbf{C}'_k$ is a column vector of constants dependent on $k$.

The boundary condition (\ref{eqn:linevapbc}) becomes \begin{equation}\label{eqn:evapstep}
v_2(x,0)=  \sqrt{\frac{2}{\pi}}\left(p_s(x,0)-\Pi(x,0)-\frac{1}{2}\theta_L(x,0)\right ) \chi_{\ell}(x),
\end{equation}
where 
\begin{align*}
\chi_\ell(x)=\begin{cases}
1, & |x|<\frac{1}{2}\\
0, & \frac{1}{2} \leq |x| \leq \frac{1}{2} + W,
\end{cases}
\end{align*}
and other moment boundary conditions remain the same, as well as the moment boundary conditions for the general $N$ moment system. The effective pressure $\Pi$ is also the generalisation for the $N$ moment system.

The remaining boundary conditions do not involve the function $\chi_\ell$, and so can be written as an infinite list of $N-1 \times N-1$ matrix boundary condition systems of the form 
\begin{align*}
\mathbf{b}_k = A_k \mathbf{C}_k,
\end{align*}
one for each $k$, similar to \eqref{eqn:bcsystem}. 
These can be solved directly to write all but one of the constants in $\mathbf{C}_k$ linearly in terms of the remaining one, leaving a single constant to be solved for, say $z_k$, for each $k$. The $z_k$ are found using \eqref{eqn:evapstep}.

The function $\chi_\ell$ can be written as a Fourier expansion, with
\begin{align*}
\chi_\ell = \sum_{j \in \Z} a_{j} e^{-i \pi j x},
\end{align*}
where
\begin{align*}
a_j = \frac{1}{2} \int_{-\frac{1}{2}}^{\frac 12} e^{i\pi j x} dx = \begin{cases}
\frac{1}{\pi j} \sin \left ( \frac{\pi j }{2}\right ) & j \neq 0 \\
\frac 12 & j=0
\end{cases}
\end{align*}
The $k$-th mode of (\ref{eqn:evapstep}) becomes
\begin{align}\label{coefbc}
v_{2,k}(0) + \sqrt{\frac{2}{\pi}}\sum_{r \in \Z} \Pi_r(0) a_{k-r} = 0  
\end{align}
The parameters $v_{2,k}$ and $\Pi_k$ are linear in $z_k$, thus we can write (\ref{coefbc}) as
\begin{align}\label{eqn:fandg}
f(k,\ell)z_k + \sqrt{\frac{2}{\pi}}\sum_{r \in \Z} g(r,\ell)a_{k-r}z_r = 0, 
\end{align}
for all $k$, for some known functions $f$ and $g$ linear in $k$.

Looking more closely at the differential equations for each mode, the zeroth mode corresponds to the one-dimensional system, so we deduce that $v_{2,0} = v_\infty$, since the continuity equation dictates the velocity is constant in $y$ for this mode. 
This means that any of the moments $\varphi^{[N]}_k$ will have at most linear dependence on $v_\infty$ and $z_n$, with no products. Therefore, the zero-th moment system will take the form
\begin{align}
\sqrt{\frac{2}{\pi}}\sum_{r \in \Z} g(r,\ell)a_{-r}z_r = -v_\infty.
\end{align}
The entire system \eqref{eqn:fandg} can be brought into the form
\begin{align}\label{eqn:infsyst}
F_{rk}z_r = -v_\infty \delta_{0k},
\end{align}
where $\delta_{ij}$ is the Kronecker delta symbol, and summation is performed over $r$. The right hand side has linear dependence on $v_\infty$, and since $F_{ij}$ has no dependence on $v_\infty$, we must have that $z_k$ is at most linear in $v_\infty$ for every $k$.
 
To verify this solution, this process is done for isothermal NSF system (with no slip or temperature jump). The matrix $Q_k$ in this case is
\begin{align}\label{incompressstokes}
Q_k = \left(\begin{matrix}
0 & i\frac{\pi k}{\ell} & 0 & \frac{-1}{\kn} \\
i\frac{\pi k}{\ell} & 0 & 0 & 0 \\
0 & -2\kn\left (\frac{\pi k}{\ell}\right )^2 & 0 & -i\frac{\pi k}{\ell}\\
-2\kn\left (\frac{\pi k}{\ell}\right )^2 & 0 & i\pi k& 0
\end{matrix}\right),
\end{align}
and then for $k\neq0$ we have
\begin{align*}
J_k = \left(
\begin{array}{cccc}
 -\frac{\pi k}{\ell}  & 1 & 0 & 0 \\
 0 & -\frac{\pi k}{\ell} & 0 & 0 \\
 0 & 0 & \frac{\pi k}{\ell}  & 1 \\
 0 & 0 & 0 & \frac{\pi k}{\ell}  \\
\end{array}
\right), && P_k = \left(
\begin{array}{cccc}
 \frac{\ell}{2 \kn \pi   k} & 0 & -\frac{\ell}{2 \kn \pi  k} & 0 \\
 -\frac{i\ell}{2\kn \pi   k} & -\frac{i\ell^2}{2 \kn \pi ^2 k^2} & -\frac{i\ell}{2 \kn \pi  k} & \frac{i\ell^2}{2 \kn \pi ^2  k^2} \\
 0 & -\frac{i\ell}{\pi  k} & 0 & -\frac{i\ell}{\pi  k} \\
 1 & 0 & 1 & 0 \\
\end{array}
\right),
\end{align*}

and
\begin{align*}
J_0 = \left(
\begin{array}{cccc}
 0 & 0 & 0 & 0 \\
 0 & 0 & 0 & 0 \\
 0 & 0 & 0 & 1 \\
 0 & 0 & 0 & 0 \\
\end{array}
\right), && P_0 =\left(
\begin{array}{cccc}
 0 & 0 & 1 & 0 \\
 0 & 1 & 0 & 0 \\
 1 & 0 & 0 & 0 \\
 0 & 0 & 0 & -\kn \\
\end{array}
\right).
\end{align*}
The $F_{jk}$ are given by
\begin{align*}
F_{jk} = \begin{cases}
\sqrt{\frac{2}{\pi}}\frac{\ell}{ \pi (j-k)} \sin \left (\frac{\pi (j-k)}{2\ell} \right )& j\neq k,\\
\frac{1}{2}\sqrt{\frac{2}{\pi}}+\frac{\ell}{2\kn \pi |j|}  & j = k \neq 0,\\
\frac{1}{2}\sqrt{\frac{2}{\pi}} & j = k = 0.
\end{cases}
\end{align*}

\begin{figure}
\includegraphics[scale=1.45, trim={0 0cm 0 0cm}]{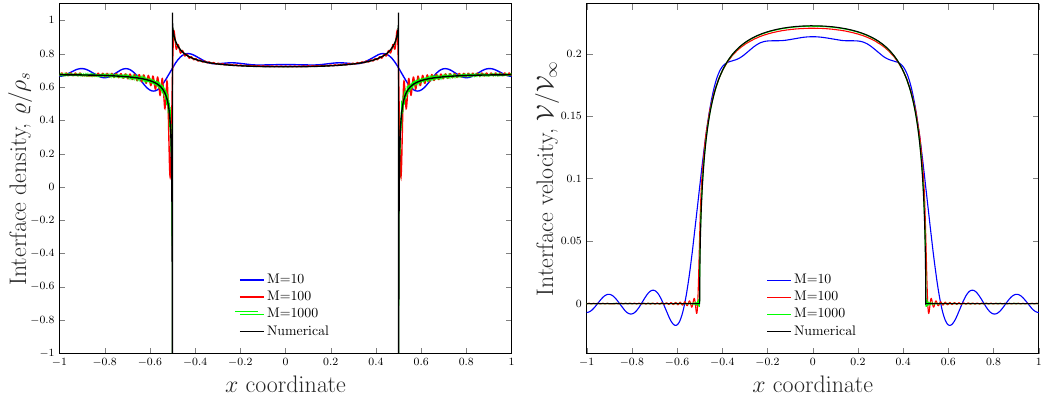}
\caption{\label{fig:fouriercomp}Velocity and density profiles along the interface for isothermal NSF, with $v_\infty=0.1$, $\phi=0.5$ (i.e. $\ell=2$), $\kn=0.05$. Comparisons between the FEM approach and the Fourier approach, showing convergence in $M$, for three values of $M$; $M=10, 100$ and $1000$.}
\end{figure}

The system (\ref{eqn:infsyst}) is solved by truncating the system, considering only $|i|\leq M$ for some $M\in \mathbb{N}$. Normal velocity and density profiles across the interface are shown in Figure \ref{fig:fouriercomp}, and are seen to converge to the numerical solution as $M$ increases.
We also see evidence of the singularity in density present in the NSF equations along the wall-liquid-vapour contact line, which we discuss in the next appendix.

\section{\label{sec:singularity}A Brief Discussion on the Singularities at the Wall-Liquid-Vapour Contact Line} 

\begin{figure}
	\includegraphics[scale=1.45, trim={0 0cm 0 0cm}]{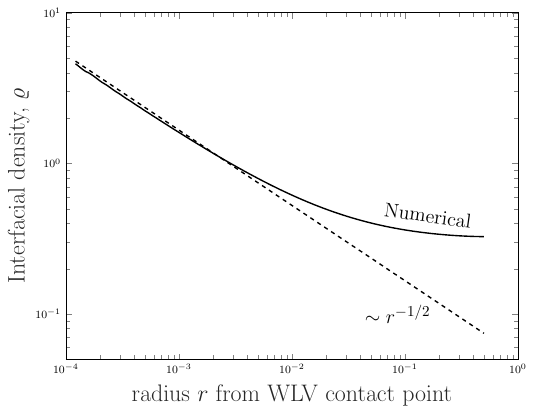}
    \caption{\label{fig:nsfsingtype}Numerically calculated density for the NSF model for increasing radius from the WLV contact point, suggesting a singularity of type $r^{-1/2}$}
\end{figure}

\begin{figure}
	\includegraphics[scale=1.45, trim={0 0cm 0 0cm}]{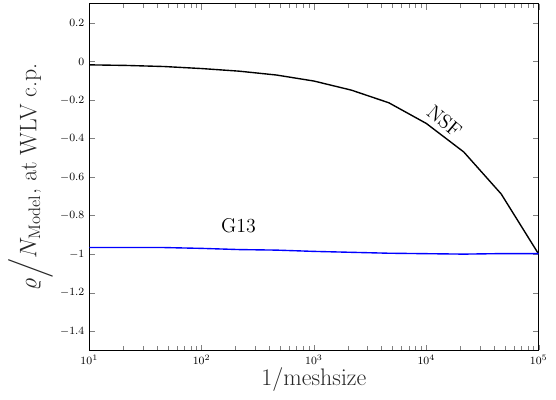}
    \caption{\label{fig:nsfvG13}Numerical values for density at the WLV contact point, for increasing mesh size, and for the NSF model and Grad 13 model. Curves are normalised with respect to the highest attained (absolute) value of density, with $N_{\text{NSF}} = 22.73$, and $N_{\text{G13}} = 0.3093$}
\end{figure}

As mentioned in the main text, the density profile seen in Figure \ref{fig:rsurfs} for NSF is caused by a singularity in the interfacial density at the Wall-Liquid-Vapour (WLV) contact point. This singularity can also be seen in Figure \ref{fig:fouriercomp} in Appendix \ref{sec:fourier}, where the density drops dramatically on the wall side of the interface.
The behaviour of pressure (and thus density) at such contact lines for Stokes flow is discussed by Nitsche and Parthasarathi \cite{nitscheseep}. For the Hertz-Knudsen-Schrage-like boundary condition 
\begin{align*}
    v_n = -p,
\end{align*}  
they found a singularity of order $r^{-\frac{1}{2}}$ in pressure, compared to the usual $r^{-1}$ type singularity found in Stokes flow under the prescription of uniform evaporative normal velocity.
Validation of this singularity class is seen in Figure \ref{fig:nsfsingtype} (recall that we assume an ideal gas, and in this instance temperature is constant, so density is proportional to pressure), where numerical results from our model are seen to be approximately proportional to $r^{-\frac{1}{2}}$ as $r\to 0$ (there will be other exponents of $r$ contributing to pressure, though $-\frac 12$ is the dominant exponent for small $r$).
While $r^{-\frac{1}{2}}$ is an integrable singularity, the singularity still manifests in the numerical model for NSF, and causes significant mesh-size-dependence for the pressure profile at the WLV contact point, as well as large spikes in values. For the singularity-free higher moment methods, finer meshes produce graphically indistinguishable curves in all figures. For instance, the relationship with mesh size is examined in Figure \ref{fig:nsfvG13} for Grad 13 and NSF. NSF is clearly dependent on the mesh size, whereas Grad 13 converges on a value as the mesh becomes finer.
For context, the mesh size at the WLV contact point for all simulations carried out in the main sections of this paper was $5 \times 10^{-3}$.

\section{\label{sec:nonlinan}Analytic Solutions for the Non-linear Boundary Condition Model in the 1D Case}
Inversion of (\ref{eqn:matrixsystem}) yields
\begin{align*}
\rho_\infty ^{[\text{G}13]} = -\frac{v_{\infty } \left(10 \pi  v_{\infty }+7 \sqrt{2 \pi }\right)}{10 \pi  v_{\infty }^2+17 \sqrt{2 \pi } v_{\infty }+16}, \quad 
\theta_\infty ^{[\text{G}13]} = -\frac{2 \sqrt{2 \pi } v_{\infty }}{10 \pi  v_{\infty }^2+17 \sqrt{2 \pi } v_{\infty }+16}, 
\end{align*}
and

\begin{align*}
\rho_\infty ^{[\text{R}13]}  &= -\frac{\sqrt{\pi } v_{\infty } \left[\left(5 \pi  \sqrt{30}+36 \sqrt{2 \pi }\right) v_{\infty } +7 \sqrt{15 \pi }+57\right]}{D},\\ 
\theta_\infty ^{[\text{R}13]} &= -\frac{2 \sqrt{\pi } v_{\infty } \left(2 \sqrt{2 \pi } v_{\infty }+\sqrt{15 \pi }+9\right)}{D}, \\
\mathbf{C}_8^{[\mathrm{R13}]} &=\frac{10 \sqrt{\pi } v_{\infty } \left(\sqrt{2 \pi } v_{\infty }+2\right)}{D},
\end{align*}
where
\begin{align*}
D = &5 \pi  \sqrt{2} \left(\sqrt{15 \pi }+6\right) v_{\infty }^2+\left(17 \pi  \sqrt{15}+115 \sqrt{\pi }\right) v_{\infty }+60 \sqrt{2}+8 \sqrt{30 \pi }.
\end{align*}
Finally, the density for the NSF system is
\begin{align*}
\rho_\infty^{[\mathrm{NSF,1D}]} = - \frac{v_\infty}{\sqrt{\frac{2}{\pi}}+v_\infty}.
\end{align*}
All expressions are seen to converge as $v_\infty \to \infty$.

\acknowledgements
This work was supported by the EPSRC (Grants No. EP/N016602/1, No. EP/P020887/1,  and No. EP/P031684/1) and an EPSRC IAA Acceleration Account (EP/R511808/1).

\bibliography{bibliography.bib}

\end{document}